\documentclass[fleqn,usenatbib]{rasti}

\usepackage{newtxtext,newtxmath}

\usepackage[T1]{fontenc}
\usepackage{multirow}
\usepackage{multicol}
\usepackage{lineno}
\usepackage{algpseudocode}
\usepackage{algorithm}
\usepackage{newtxtext,newtxmath}
\usepackage{tikz}
\usetikzlibrary{fit}
\usetikzlibrary{shapes.geometric,arrows}
\usepackage{svg}
\usepackage{caption}
\usepackage{subcaption}

\DeclareRobustCommand{\VAN}[3]{#2}
\let\VANthebibliography\thebibliography
\def\thebibliography{\DeclareRobustCommand{\VAN}[3]{##3}\VANthebibliography}

\usepackage{graphicx}	
\usepackage{amsmath}

\newcommand{\profound}{\texttt{ProFound}}
\newcommand{\pybdsf}{\texttt{PyBDSF}} 
\newcommand{\DRUID}{\texttt{DRUID}}

\newcommand{\Druid}{\DRUID}
\newcommand{\new}{} 

\setcitestyle{authoryear, open={((},close={)}}

\title[\DRUID: Source Detection and Deblending in Astronomical Images with Persistent Homology]{\DRUID: Source Detection and Deblending in Astronomical Images with Persistent Homology}

\author[R. A. Shaw et al.]{
R. A. Shaw$^{1}$\thanks{E-mail: rhys.shaw@bristol.ac.uk},
S. Fotopoulou$^{1}$,
M. Birkinshaw$^{1}$,
N. Maddox$^{1}$,
H. Stewart$^{1}
$
\\
$^{1}$School of Physics, HH Wills Physics Laboratory, Tyndall Ave, Bristol BS8 1TL \\
}

\date{Accepted XXX. Received YYY; in original form ZZZ}

\pubyear{2024}

\begin{document}
\label{firstpage}
\pagerange{\pageref{firstpage}--\pageref{lastpage}}
\maketitle

\begin{abstract}
Source detection is a vital part of any astronomical survey analysis pipeline. 
In addition, a versatile source finder that can recover and handle sources of all morphological types is becoming more important as surveys get bigger and achieve a higher resolution than ever before.
Here we present Detector of astRonomical soUrces in optIcal and raDio images (\DRUID), a source finder that utilises persistent homology to detect and deblend sources. 
This method enables us to effectively and uniquely segment structures within \new{morphologically} complex sources and deal with high source density images. 
We test \DRUID\ on the complex morphologies of 3CR radio loud active galactic nuclei, where we \new{demonstrate its} ability to usefully segment the main structures in the sources. 
We also demonstrate the level of structure \DRUID\ segments within well resolved galaxies \new{in the optical}.
\new{Finally, we present two source catalogues on the LoTSS Deep field observation of the Lockman Hole} and an example tile from the KiDS r-band survey. 
We conclude that \DRUID 's method \new{of} utilising persistent homology provides a new way to detect and deblend highly nested sources.

\end{abstract}





\section{Introduction}

Source detection software has been used in astronomy for a significant amount of time to automatically generate catalogues of galaxies \new{--} an example is FOCAS \citep{jarvis_focas_1981}. This early software led to the development of SExtractor \citep{bertin_sextractor_1996}, a source finder known for its robustness and effectiveness in processing large astronomical images, and its latest variant SExtractor++ \citep{bertin_sourcextractor_2020}. SExtractor has been widely used to generate catalogues of stars and galaxies from a range of astronomical surveys. However, it is not suited to crowded fields or complex morphologies such as radio lobes. With the development of new instruments and larger surveys, going deeper into the sky with a higher resolution than ever before, source finders must perform well with the influx of more data with dense and complex images.

Within the Radio domain, the construction of the Square Kilometre Array (SKA) is underway, and we will soon be able to observe galaxies down to the nJy scale between \new{50 MHz < $\nu$ < 15.4 GHz}. Some of its precursor observatories have already done this, including the Meer Karoo Array Telescope \citep[MeerKAT;][]{booth_meerkat_2009}, the SKA MID precursor, the Australian Square Kilometre Array Pathfinder \cite[ASKAP;][]{hotan_australian_2021} which will observe mid-frequency GHz emission. The low-frequency end is being observed with the Low Frequency Array \cite[LOFAR;][]{van_haarlem_lofar_2013} and the Murchison Widefield Array  \citep[MWA;][]{tingay_murchison_2013}, the SKA low-frequency precursor. This is also complemented by the upgrading of existing facilities such as the Karl G Jansky Very Large Array \citep[VLA;][]{thompson_very_1980}. These \new{i}nstruments when fully operational will allow for the detection of millions more radio sources creating larger populations of sources at deeper sensitivities than ever before. This reaffirms the need for a source finder and techniques that can meet the challenges of this new frontier in Astronomy.

Typically, in radio surveys of the extragalactic sky the source populations fit into two categories. \new{The first} category consists of star-forming galaxies (SFGs) whose emission originates from \new{electrons accelerated by} supernova remnants that emit mainly \new{through} synchrotron at GHz frequencies \citep{condon_radio_1992}. \new{The second category comprises} Radio Loud Active Galactic Nuclei (RLAGN) that \new{emit through synchrotron emission, which is produced by relativistic electrons interacting with magnetic fields in the jets and lobes associated with the accreting supermassive black hole}. The radio emission from supernovae remnants is then an indicator \new{of} the star formation rate of the host galaxy \citep{smith_lofar_2021}. RLAGN have a complex and varied morphology, \cite{fanaroff_morphology_1974} defined two Classes that well represent \new{the} main variation in RLAGN morphology. An FRI \new{has} jets that merge into diffuse plumes and an FRII contains jets that lead to hotspots and lobes \citep{hardcastle_radio_2020}. The mechanism that generates this emission is also thought to be variable, leading to fading emission remnants that we observe in the radio sky. An increasing field of view, resolution and sensitivity to low surface brightness features will make these morphology types more frequent in radio surveys.

Source detection in radio surveys is commonly done by fitting Gaussians to an island of flux above the noise to extract flux densities. 
\new{This relies on estimating the RMS noise to ensure proper source extraction.}
Current software like \pybdsf\ \citep{mohan_pybdsf_2015} and \texttt{AEGEAN} \citep{hancock_compact_2012,hancock_source_2018} fit Gaussian components to flux islands to generate a catalogue of components and sources. This methodology works well with point and compact sources, where they \new{are usually fitted with one or a small number of Gaussians}. However, well-resolved and extended sources like complicated jet morphologies will be fitted with multiple Gaussians that are combined to create a single source. Modelling the emission from RLAGN FRI, FRII and large SFGs with Gaussians do not necessarily work well as the morphologies become increasingly complex. This highlights the need for alternative methods for extracting flux densities of these more complex objects. Flux density measurements that are accurate and complete are crucial to many measures of the extra-galactic sky. This includes source counts of radio populations which can be used to check and compare cosmological models \citep{condon_resolving_2012, hale_mightee_2023}. It is also important \new{i}n calculating radio luminosity functions \citep{prescott_galaxy_2016,smolcic_vla-cosmos_2017} and calculating spectral indices and modelling spectral sources like supernova remnants \citep{callingham_low_2016,balzan_radio_2022}.

Within the optical and infrared regime, popular source finders include the previously mentioned SExtractor \citep{bertin_sextractor_1996} and \profound \citep{robotham_profound_2018}. Source extraction generally begins with creating a model of the background sky. This can be a complicated task when there is a significant gradient in the sky, caused by several effects including bad data reduction and contamination from large extended objects in the line of sight. Additionally, background \new{estimation} can be further complicated when images become confused. With the background sky model, we can subtract it from our image leaving islands of flux that we call sources. The treatment from this point onwards varies significantly and might need to be tailored for the requirements of the \new{study}. 
\new{\pybdsf\ does distinguish between sources and components and will group components based on user-defined parameters.}
Islands of flux can be further separated into separate sources or components of a source; this is usually referred to as deblending. Finally, properties of the sources are found, and a catalogue is created. 
Catalogues also require some level of component association. This is a particular problem within the radio regime as Gaussian components from \pybdsf\ tend to require extensive manual association to correctly match components into sources using multi-wavelength counterparts \citep{whittam_mightee_2024,hardcastle_lofar_2023,williams_lofar_2019}.

When a detection is made through thresholding an image, we must check if there is more than one source present. In a simple case, this can be done by analysing the number of peaks with the number of troughs within the Island but is easily complicated with the addition of noise. As previously mentioned \pybdsf\ deals with deblending by fitting the observed emission with a sum of Gaussian components, thereby creating two or more sources with modelled emission. \profound\ does this in an alternative fashion where peaks of flux above a specified threshold are kept for the watershed dilation process. These final components are grouped into sources, based on the sharing of flux islands. The assignment of pixels to source regions leads to morphologically unbiased detection and measurements of fluxes. This allows for the inclusion of otherwise missing fluxes from more extended regions of the source. At optical and infrared wavelengths this is advantageous due to \new{the} rich morphology types of galaxies, which can consist of a combination of bars, discs, and spiral arms. SExtractors method for deblending consists of doing recursive thresholding to generate a tree diagram of peaks and troughs in the image and the\new{n} deciding which are sources based on their relative brightness. SExtractor then evaluates the flux from the source using fitted or fixed apertures. \cite{haigh_optimising_2021} investigated the optimised performance of a range of source finding methods, including SExtractor and \profound, in a range of situations. They concluded that despite each source finder having a reasonable level of detection completeness a shared weakness was in the ability to accurately deblend nested objects.

In this work we present a new technique for identifying potential sources within astronomical images, using persistent homology. We use persistent homology to both find the sources and handle complex morphologies of both RLAGN and well resolved galaxies. \new{In} Section \ref{sec:persistent_homology} we give an overview of persistent homology and how it can be applied to the analysis of astronomical images. Section \ref{sec:Methods} presents the new source finder \DRUID\ and the structure and features of the program. \new{In} Section \ref{sec:simulations} we test \DRUID\ on simulated point sources injected into an r-band image from the Kilo Degree Survey (KiDS) \citep{wright_kidsviking-450_2019} and evaluate its ability to recreate the injected source catalogue. \new{In} Section \ref{sec:Analysis_3c} we investigate how \DRUID\ performs when handling the complex morphology of RLAGN from the 3CR catalogue, and we compare it to other source finders. In Section \ref{sec:LoFarLockmanHole} we create a source catalogue of \new{LOFARs LoTTS Deep Field observation} of the Lockman hole \citep{tasse_lofar_2021} and compare it to the \pybdsf\ derived catalogue \new{along with the optical crossmatched catalogue \citep{kondapally_lofar_2021}}. Finally, in Section \ref{sec:KiDs} we create a catalogue of sources from an r-band image from the KIDS survey and compare the outcomes of our runs of \profound\ against the SExtractor-derived associated catalogue.







\begin{figure*}
    \centering
    \begin{subfigure}[b]{0.6\textwidth}
        \includegraphics[width=\textwidth]{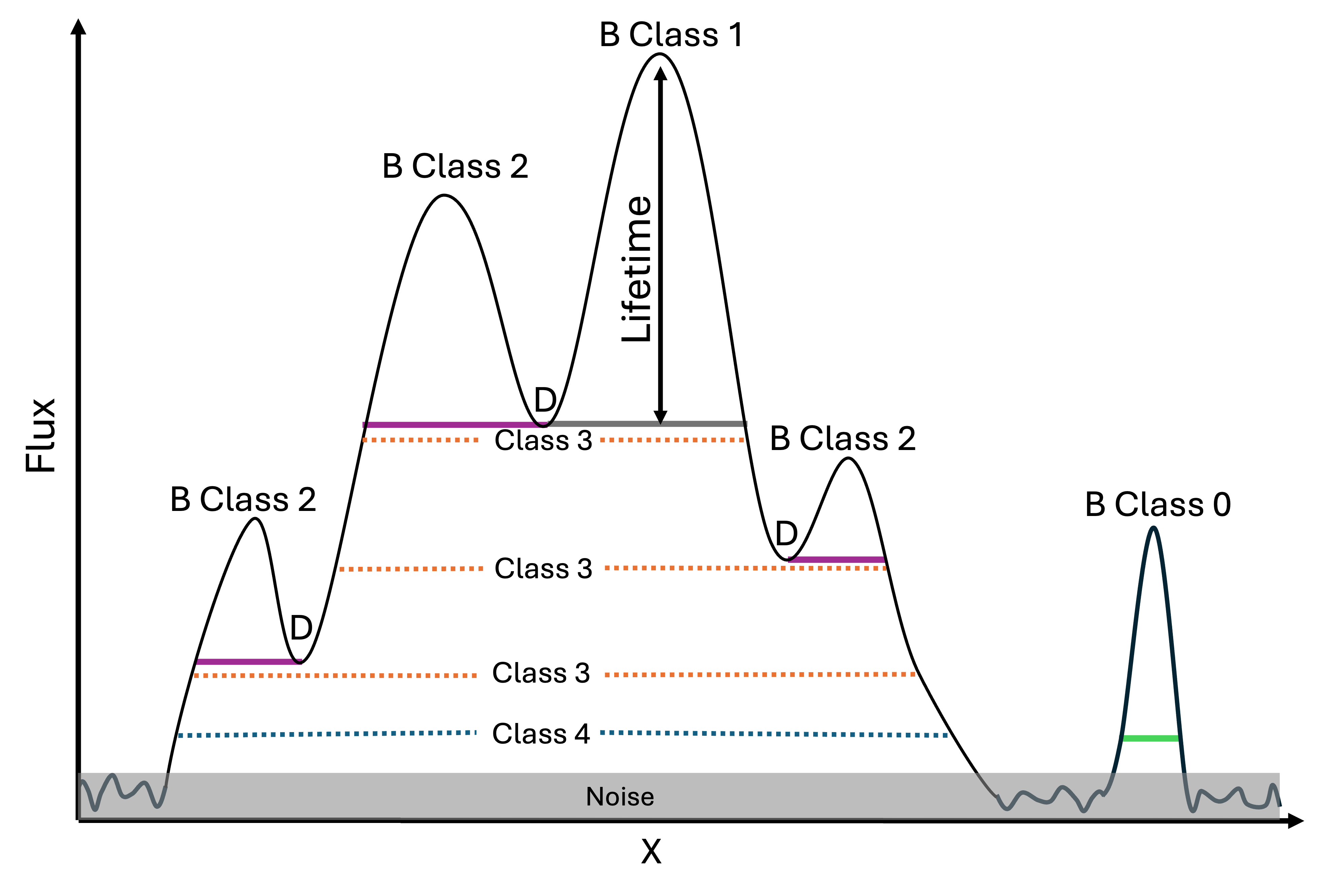}
    \end{subfigure}
    \begin{subfigure}[b]{0.35\textwidth}
        \includegraphics[width=\textwidth]{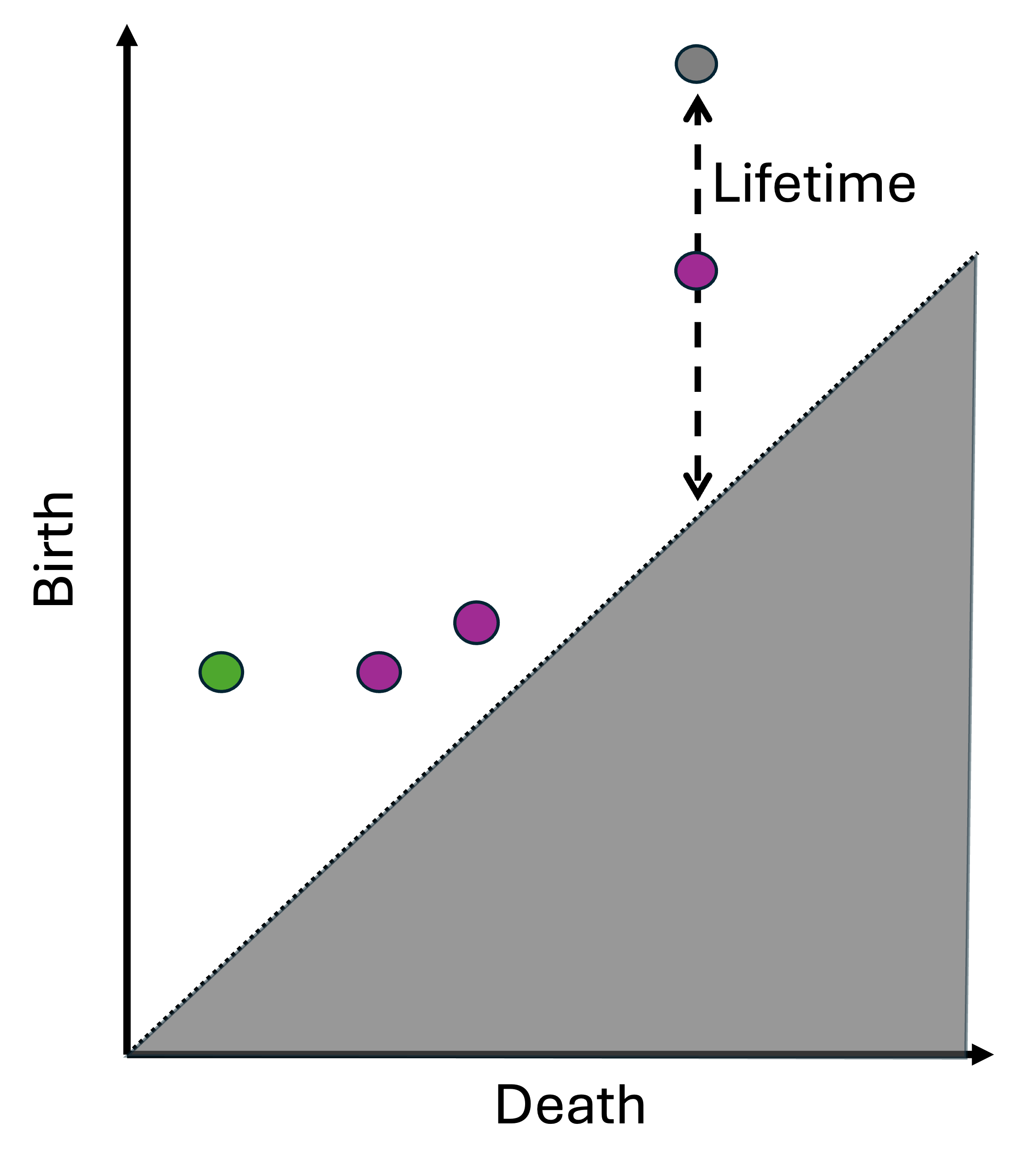}
    \end{subfigure}
    \caption{\new{On the left, we show the idealised cross-section of a flux Island with annotations of where DRUID finds Birth (B) and Death (D) locations along with the contours that are drawn as a result. The Class of each component is labelled above the flux distribution for Classes 0, 1 and 2. Classes 3 and 4 always contain other Classes, they are marked within the Island with a dashed line. Class 3 is defined at the merging location of two or more components and Class 4 marks the adopted detection threshold of the image. A single point source is labelled as Class 0 and its line is drawn at the detection threshold of the image. The persistence diagram that would be calculated from this image is shown on the right. Annotated on both plots is the lifetime of the brightest component.}}
    \label{fig:drawn_example_of_DRUID}
\end{figure*}

\section{Persistent Homology}
\label{sec:persistent_homology}

Persistent homology is a method that computes the topological features of a space at varying resolutions. 
It provides a way to quantify the connectivity between features within that space and is a powerful tool to analyse the presence of these features within data \citep{hensel_survey_2021}.

\subsection{Complexes}

The first step of persistent homology is the construction of complexes. Complexes provide a classical approach to representing discrete objects. They consist of interconnected geometric shapes called simplices. The dimensions of these simplices are categorised as follows: a 0-simplex represents a single point, a 1-simplex is an edge, 2-simplex forms a square, 3-simplex constructs a cube, and so on. The dimension of a simplex $\sigma$, is equivalent to its intrinsic dimension. This dimension arises from the number of vertices minus one. Within a simplex $\sigma$, there are faces, which are smaller simplices. A complex $\Sigma$ is a collection of simplices that adhere to specific rules. First, all faces of a simplex in $\Sigma$ are themselves part of $\Sigma$. Second, whenever two simplices in $\Sigma$ intersect, the resulting common region is a face of both simplices. The dimension of a simplicial complex is the highest dimension among its constituent simplices. This value reflects the dimension of the largest simplex within the complex.

\subsection{Persistent Homology}

\begin{figure*}
    \centering
    \begin{subfigure}[b]{0.36\textwidth}
        \includegraphics[width=\textwidth]{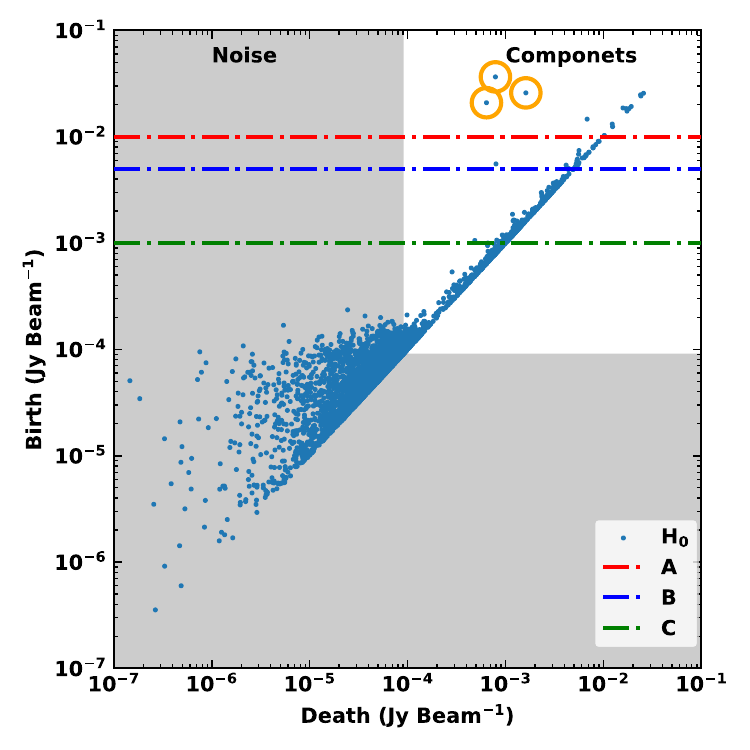}
        \caption{Persistence Diagram.}
    \end{subfigure}
    \begin{subfigure}[b]{0.33\textwidth}
        \includegraphics[width=\textwidth]{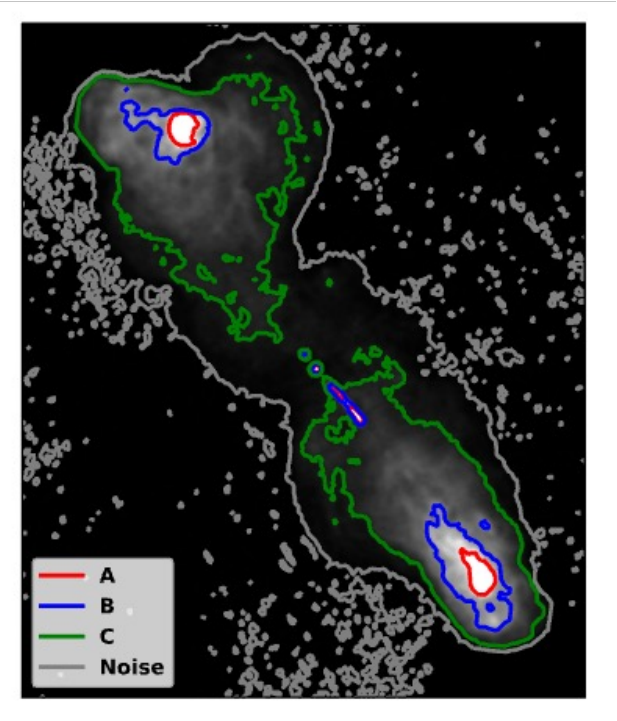}
        \caption{3C219.}
    \end{subfigure}
    \begin{subfigure}[b]{0.3\textwidth}
        \begin{subfigure}[b]{\textwidth}
            \includegraphics[width=\textwidth]{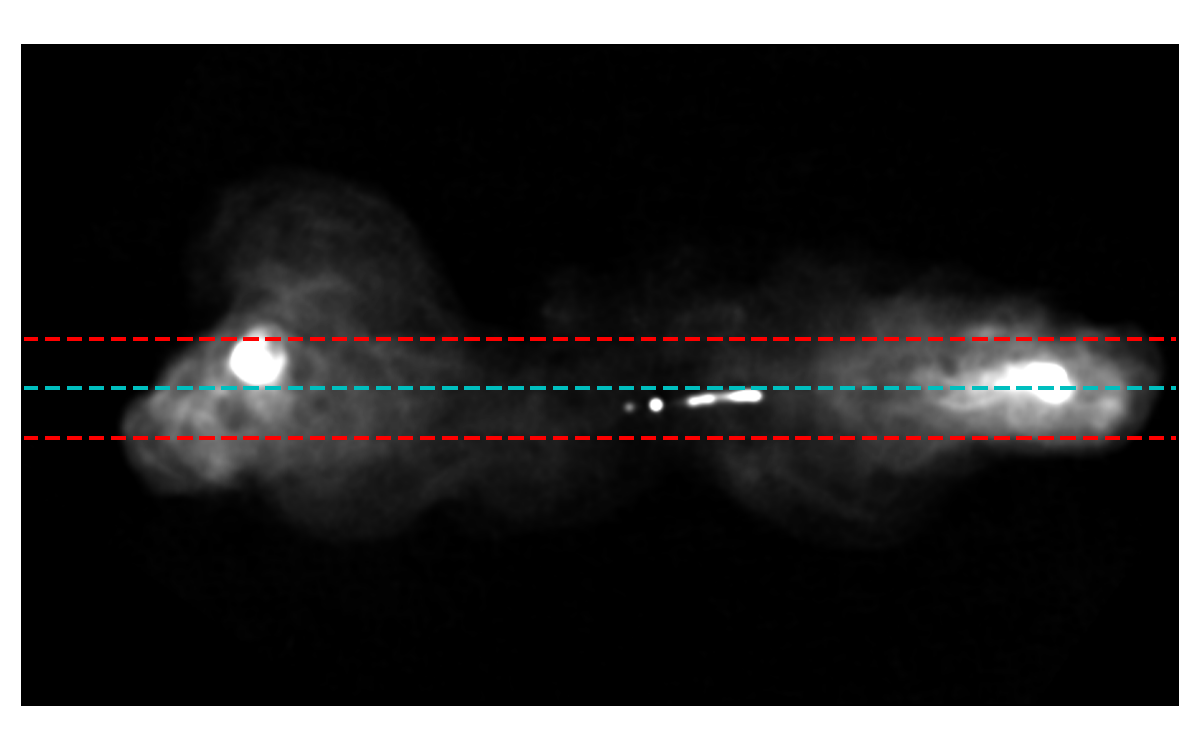}
            \caption{3C219.}
        \end{subfigure}
        \begin{subfigure}[b]{\textwidth}
            \includegraphics[width=\textwidth]{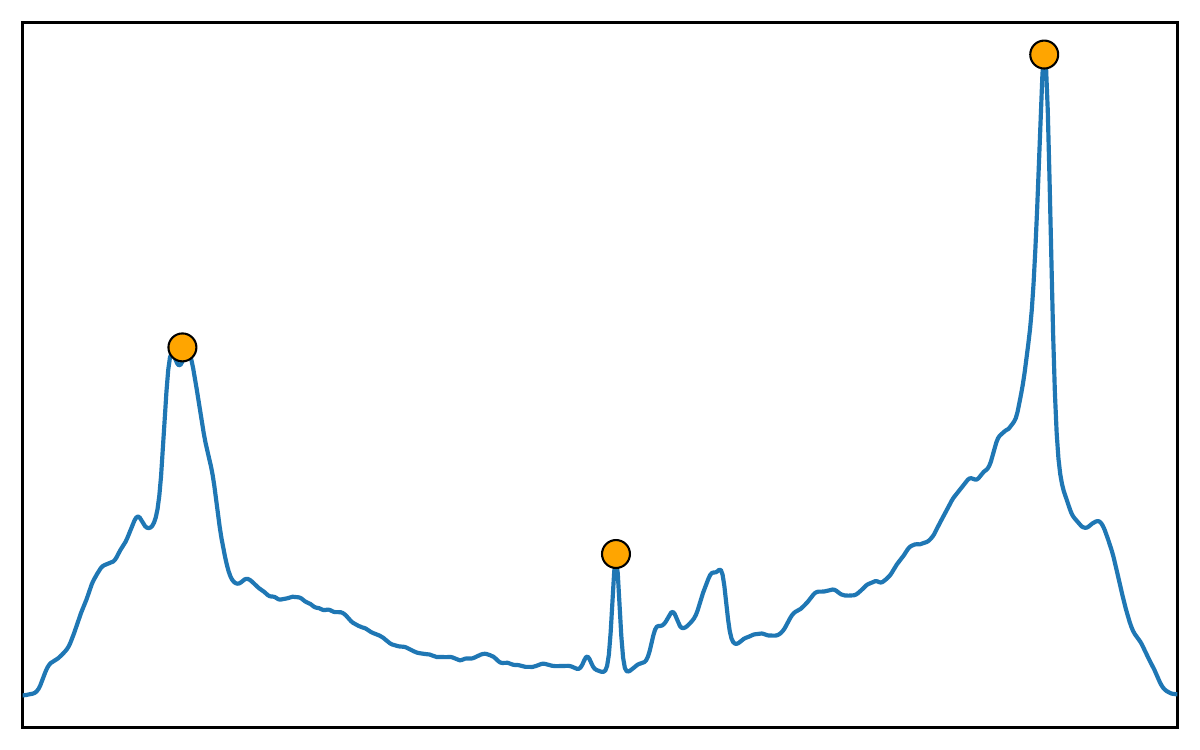}
            \caption{3C219 Cross-section.}
        \end{subfigure}
    \end{subfigure}
    \caption{\new{Persistence diagram (a) with the reference image of 3C219 \citep{clarke_origin_1992} (b). The lines A, B and C show three threshold flux levels. The contours on the reference image (b) are drawn at the same flux levels as their corresponding coloured lines on the persistence diagram. The quadrants labelled as noise are below the measured Median Absolute Deviation of the noise from the reference image. Plot (d) shows a cross-section of the source created though averaging the pixels between the red lines shown in (c). This cross-section demonstrates the three prominent peaks on the image that correspond to the circled orange points on diagram (a).}}
    \label{fig:pd_example}
\end{figure*}

Homology groups are algebraic structures that describe cycles and boundaries in the space. To define homology groups, \new{we} start with a complex composed of a set of simplices. Chains are formed from linear combinations of simplices, these chains represent cycles or loops within the complex. Homology groups are ranked by their Betti number \new{--} the rank of each group $H_{n}$ is called the n-th Betti number. The 0-homology group $H_{0}$ represents path connected components (islands). The 1-homology ($H_{1}$) represents a one-dimensional hole. $H_{0}$ can simply be distinct continuous objects that are not touching each other. $H_{1}$ holes are holes in a 2D object like a ring. A space can be composed of multiple types of homology groups. A key part of homology is \new{the type of complex used} to construct homology groups, for this application, we will be using cubical complexes. For grid structured data, like an image, using a cubical complex is the most natural. A cubical complex is constructed at the intersection of pixels, edges, and neighbours \new{and it} consists of a set of points, line segments, squares, and cubes. 

Persistent homology measures the persistence of the topological features (homology groups) across different scales. 
\new{In image analysis, filtration involves examining pixels from highest to lowest intensity. As lower intensity pixels are progressively incorporated, homology groups are constructed}.
\new{Persistent homology offers a distinct approach to image analysis compared to traditional methods like thresholding, peak detection, and watershedding. It examines the evolution of topological features across multiple scales, tracking how features appear, persist, and disappear as thresholds change. This multiscale approach provides a view of local and global image structures, quantifying feature significance based on persistence.}
The \new{persistence} information is usually displayed as a persistence diagram, where each homology group is described by the places where these complexes meet (point of its death) and where it begins (its birth location) along with the pixel value at this location. The \new{b}irth locations are equivalent to the location of the brightest pixel in that complex and the death location is the same as the trough where two complexes meet.

Figure \ref{fig:drawn_example_of_DRUID} shows a cross-section of \new{an idealised flux Island:} the filtration process constructs homology groups as more of the image at lower flux is revealed. \new{This is accompanied by the corresponding persistence diagram}. The birth and death locations are labelled by B or D, \new{respectively}, \new{along} with the \DRUID\ Class it belongs to (further discussed in Section \ref{sec:code_diagram}). The lifetime in Figure \ref{fig:drawn_example_of_DRUID} represents the difference in pixel value between the birth and death locations \new{and is the same as the lifetime marked on the persistence diagram.}

A demonstration of filtration of a radio image with the associated persistence diagram is shown in Figure \ref{fig:pd_example}. \new{The thresholds A, B and C show the regions of the image that are incorporated into complexes during the filtration. Three points are highlighted in orange which correspond to the complexes created at each of the peaks in the cross-section plot (right). The  grey-shaded part of the diagram is not useful for source detection as it is below the noise threshold.} 
As shown in Figures \ref{fig:drawn_example_of_DRUID} and \ref{fig:pd_example} persistence diagrams provide important information \new{that can be used} for segmenting the image, invariant to deformations like rotations, transformations, or image warping.
The source’s \new{extended shape} does not affect its persistent homology. 
The \new{persistence diagram} gives information about the features that persist for a long range of values across the filtration. During the construction of persistence diagrams, an elder rule is followed. If two of the same type of features are merged, then by convention the youngest of them (the one with the most recent birth time) dies. This results in the brightest pixel having an infinite persistence. This persistence diagram contains information about where, for all scales, peaks merge in the image, and this can be used to automatically assign boundaries between them. Additional details and a more technical description of persistent homology can be found in \citep{hensel_survey_2021,chazal_introduction_2021}.




\section{Methods}
\label{sec:Methods}

\subsection{\Druid}
\label{sec:DRUID}
\begin{table}
\begin{tabular}{ll}
\hline
Parameter   & Description                     \\ \hline
ID          & Unique source number            \\
Birth       & Value of pixel at Birth location\\
Death       & Value of pixel at Death location\\
RA          & Right Ascension (J2000)         \\
DEC         & Declination (J2000)             \\
Xc          & Flux weighted centroid  X       \\
Yc          & Flux weighted centroid  Y       \\
x1          & x position of the brightest pixel (Birth location) \\
y1          & y position of the brightest pixel (Birth location) \\
x2          & x position of closest saddle point (Death location) \\
y2          & y position of closest saddle point (Death location) \\
Maj         & Polygon boundary major axis     \\
Min         & Polygon boundary minor axis     \\
PA          & Position angle of polygon fitted ellipse\\
Area         & Number of pixels in boundary    \\
Flux\_total  & Flux from the source            \\
Flux\_err    & Error of Flux\_total (for optical/CCD images)\\
Flux\_peak   & Peak flux from the source      \\
CorrF        & Beam/PSF correction factor     \\
Class        & The Class of the source         \\
parent\_tag  & ID of the parent source        \\ 
enclosed\_i  & List of IDs that this source is a parent of \\
edge\_flag   & Flag indicator of location at edge of image. \\ \hline
\end{tabular}
\caption{\new{List of \DRUID's key output parameters.}}
\label{tab:DRUID_outputs}
\end{table}

The Detector of astRonomical soUrces in optIcal and raDio images (\DRUID)\footnote{\href{https://github.com/RhysAlfShaw/DRUID}{https://github.com/RhysAlfShaw/DRUID}.} was designed to be easy to use and allow for customisation of \new{the analysis pipeline} to remain flexible to changing needs, whilst including robust default processing techniques. Python was \new{the chosen development language} for two main reasons; it is widely used within the astronomy community and important packages exist that \DRUID\ is dependent on to calculate the persistence diagrams. \DRUID\ utilises the results from persistent homology, discussed in Section \ref{sec:persistent_homology} on an image to create sources and components of sources. \DRUID\ has been designed to work on both optical and radio images and consequently has two operating modes to deal with the nuances of each domain. \new{Large images are handled by creating sub-images of a specified size, with an overlapping buffer which can also be specified. Duplicates from the overlap regions are removed in favour of the source that is closest to the centre of its sub-image.}


\tikzstyle{startstop} = [rectangle, rounded corners, minimum width=3cm, minimum height=1cm,text centered, draw=black, fill=red!30]

\tikzstyle{io} = [trapezium, trapezium left angle=70, trapezium right angle=110, minimum width=3cm, minimum height=1cm, text centered, draw=black, fill=blue!30]

\tikzstyle{process} = [rectangle, minimum width=3cm, minimum height=1cm, text centered, draw=black, fill=orange!30]
\tikzstyle{decision} = [diamond, minimum width=3cm, minimum height=1cm, text centered, draw=black, fill=green!30]

\tikzstyle{arrow} = [thick,->,>=stealth]

\tikzstyle{startstop} = [rectangle, rounded corners, 
minimum width=2cm, 
minimum height=1cm,
text centered, 
draw=black, 
fill=red!30]

\tikzstyle{io} = [trapezium, 
trapezium stretches=true, 
trapezium left angle=70, 
trapezium right angle=110, 
minimum width=1cm, 
minimum height=1cm, text centered, 
draw=black, fill=blue!30]

\tikzstyle{process} = [rectangle, 
minimum width=1cm, 
minimum height=1.5cm, 
text centered, 
text width=3cm, 
draw=black, 
fill=orange!30]

\tikzstyle{decision} = [diamond, 
minimum width=1cm, 
minimum height=1cm, 
text centered, 
draw=black, 
fill=green!30]

\tikzstyle{merger} = [circle,
    draw=black,
    fill=green!30]

\tikzstyle{arrow} = [thick,->,>=stealth]
\tikzstyle{line} = [thick]

\tikzstyle{mycomment}=[inner sep=1pt,scale=0.75,align=left]


\begin{figure*}
    \centering
    \begin{tikzpicture}[node distance=2cm]
    
    \node (in1) [io] { Input params};

    \node (dec1) [decision, right of=in1, xshift=1.5cm] {if \texttt{cutup}};
    \node (label1) [mycomment, right of=dec1, xshift=0.5cm] {True};
    \node (label2) [mycomment, below of=dec1, xshift=4cm, yshift=0cm] {False};
    \node (pro1) [process, right of=dec1, xshift=2cm] {Generate sub-images};
    
    \node (pro2) [process, right of=pro1, xshift=+3cm] {Create background map};
    \node [draw,black,thick, inner sep=5pt, fit=(pro1) (dec1) (label1) (label2)] (box1) {};
    \node (name) [above of=pro1,xshift=-1cm, yshift=-0.75cm]{\texttt{sf.\_\_init\_\_()}};
    \draw [arrow] (in1.east) -- (dec1.west);
    \draw [line] (dec1.south) |- (label2);
    \draw [line] (dec1.east) -- (label1);
    \draw[arrow] (label2.east) -| (+10,0);
    \draw[arrow] (label1.east) -- (pro1.west);
    \draw[arrow]  (pro1.east) -- (pro2.west);

    \node (dec2) [decision, below of=pro2, yshift=-3cm] {Sub-image processed?};
    \node (label3) [mycomment, below of=dec2, yshift=-0.5cm] {False};
    \node (label4) [mycomment, above of=dec2, yshift=+0.5cm] {True};
    \node (pro3) [process, below of=label3,yshift=+0.5cm] {Generate persistence diagram [\texttt{cripser}]};
    \node (pro4) [process, below of=pro3, yshift=-0.5cm] {Remove below detection threshold};
    \node (pro5) [process, left of=pro4, xshift=-2cm] {Flag edge sources};
    \node (pro6) [process, above of=pro5, yshift=0.5cm] {Calculate area and bbox};

    \draw[arrow] (pro2.east) -- (+15,0) |- (dec2.east);
    \draw[line] (dec2.north) -- (label4.south);
    \draw[arrow] (pro6.north) |- (dec2.west);
    \draw[line] (dec2.south) -- (label3.north);
    \draw[arrow] (label3.south) -- (pro3.north);
    \draw[arrow] (pro3.south) -- (pro4.north);
    \draw[arrow] (pro4.west) -- (pro5.east);
    \draw[arrow] (pro5.north) -- (pro6.south);

    \node (pro7) [process, left of=label4, xshift=-6cm,yshift=-1cm] {Remove duplicates};
    \node (pro8) [process, left of=pro7,xshift=-2cm] {Create hierarchy};
    \node (pro9) [process, below of=pro8,yshift=-0.2cm] {Correct First destruction (CFD)};
    \node (pro10) [process, right of=pro9, xshift=2cm] {Assign \texttt{parent\_ID} and classify};

    \draw [black, thick, label={phsf()}] (-2,-2.5) -- (+14.5,-2.5) -- (+14.5,-12) -- (+6,-12) -- (+6, -7.5) -- (-2,-7.5) -- (-2,-2.5);
    \node (labelname) [below of=pro1, yshift=-0.25cm]{\texttt{sf.phsf()}};
    \draw [arrow] (label4.west) -| (pro7.north);
    \draw [arrow] (pro7.west) -- (pro8.east);
    \draw [arrow] (pro8.south) -- (pro9.north);
    \draw [arrow] (pro9.east) -- (pro10.west);

    \node (pro11) [process, below of=pro10, yshift=-0.5cm,xshift=-2cm] {Source characterisation};
    \node (end1) [io, below of=pro11] {Object returned};

    \draw [arrow] (pro10.south) |- (pro11.east);
    \draw [arrow] (pro11.south) -- (end1.north);

    \end{tikzpicture}
    
    \caption{Detailed code diagram for \DRUID, this is separated into two main components the initialisation when the input parameters are parsed, followed by the persistent homology source finding function that preforms the persistent homology source finding and deblending. The code ends with source characterisation and the object is returned.}
    \label{diagram:DRUID_FlowChart}
\end{figure*}
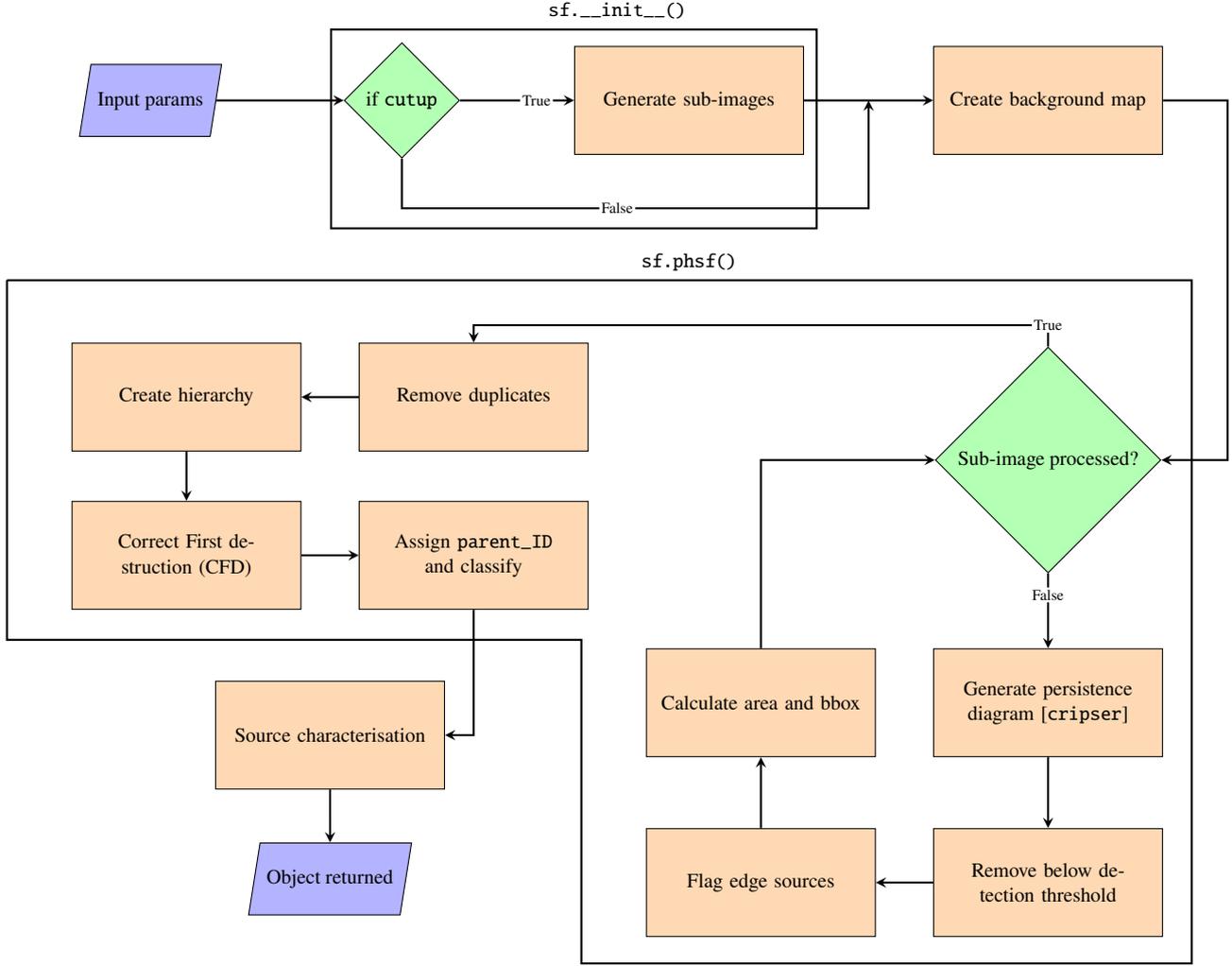


\DRUID\ bases the core of the source finding on the persistent homology analysis that is carried out by the \texttt{cripser} python package \citep{kaji_cubical_2020}. This is a cubical implementation of the popular persistent homology package \texttt{ripser} \citep{Bauer2021Ripser}. It also offers one of the fastest computations of the persistence diagram of a NxN array. The rest of the package works to post-process this information to generate segmentations of sources and relationships between the detected components. Finally, \new{this is} followed by the determination of source properties and construction of a catalogue of detected sources.

\subsubsection{Code Diagram}
\label{sec:code_diagram}
A detailed code diagram of \DRUID\ is shown in Figure \ref{diagram:DRUID_FlowChart}. \DRUID\ is made up of two main parts.  Initialisation, where the input image and parameters are set to process the image ready for its persistent homology to be determined. Secondly, \new{the} \new{p}ersistent homology \new{section} which calculates the persistence diagram using \texttt{cripser}, removes noise, and performs some basic segmentation to create relationships between components. This is followed by the calculation of source properties. \new{In more detail:}

\begin{enumerate}

    \item Initialisation (\texttt{sf.\_\_init\_\_()}) - Here global variables for the Python class are set including image or image path, whether to apply smoothing, limits on the minimum area of the source and mode (working with radio/optical) which affects the processes done when called later. It is also where \DRUID\ will \new{create sub-images} with a defined overlapping buffer. Additionally, the option to send some matrix operations to the GPU is available here. This acceleration significantly improves the performance when working with larger arrays (above 1000x1000 pixels).

    \item Background Estimation - Before we analyse the persistence diagram computed on the image, we need to evaluate the background so that we can remove points from the persistence diagram that are below the noise and ensure correct parent association later. The background can be calculated in several ways this includes root mean square ($RMS$), median absolute deviation ($\sigma_{MAD}$) and SExtractor sigma clipping. For a set of $n$ values $X = {x_{1},x_{2},...,x_{n}}$, the $RMS$ is expressed as
    \begin{equation}
        RMS = \sqrt{\frac{1}{n} (x^{2}_{1} + x^{2}_{2} + ... + x^{2}_{n})},
    \end{equation}
     and the $\sigma_{MAD}$ is expressed as,
     \begin{equation}
         \sigma_{MAD} = median(|x_{i} - median(X)|).
     \end{equation}
     
     Sigma clipping involves sampling several pixels in the image\new{,} measuring the properties of this distribution and iteratively removing pixels beyond $3\sigma$ until it reaches convergence. Any of these methods can be used to create a background map with a specified \new{sliding box size}, similar to \pybdsf. It is important in the radio regime to include this functionality as it can reduce the number of spurious detections around a bright source with significant artefacts. To reduce this type of spurious detections the box size should be set to the scale over which these artefacts are present. We include the estimation with $\sigma_{MAD}$ as we have found it to occasionally be \new{favoured over} the $RMS$, especially \new{for images} with \new{high source density} and high dynamic range. \new{The background is calculated from the original image, not from sub-images created in the initialisation.}

    \item Calculating and processing the persistence diagram (\texttt{sf.phsf()}) - For each \new{sub-}image, we calculate the persistence diagram handled by \texttt{cripser}. After this has been computed we remove any points with \new{b}irths that are below the detection threshold, this marks the boundary between features in the image and noise. Next, we set any death point less than the specified analysis threshold to \new{the analysis threshold} value. This sets a lower limit on pixels that are included in later segmentation. This is followed by the removal of components less than the specified area limit, which is important to remove spurious detection due to small fluctuations in the background around bright sources. \new{The processing of the sub-images is the most computationally demanding, therefore we provide GPU acceleration for this part}. \new{After all sub-images have been processed, duplicates from within overlapping regions are removed. This is done by assigning an edge flag when part of the source mask is at the sub-image edge or within the buffer zone. Duplicates are matched on their birth locations and value, and the one closest to its sub-image centre is kept.}
    \new{This is} followed by the making of parent-child associations between regions, \new{and it} allows us to have a hierarchy of components identifying islands of connected emission. Finally, we correct for the first destruction (CFD process) of the brightest component in the Island. This is important as otherwise the extended emission is automatically associated with the brightest pixel’s component. Because of persistent homology, the brightest pixel will never die as its complex will consume the entire image. Without correcting for this, we will lose any characterisation of the brightest component when it first destroys a neighbouring complex. This is followed by labelling the \texttt{parent\_id} and the class of the component. \new{The class of a source is assigned with Algorithm \ref{algorithm:DRUID_Classes}. If a component contains no children and has no parent, i.e. it is both not inside another component and does not itself contain one, it is assigned class 0. When it does not have any children but is inside of another component this is assigned Class 2. If it does contain children and has a parent this is the intermediate class 3. Then if the component has no parent but contains children this is class 4. Class 1 is then assigned if the component was created from the CFD process and must then be the highest peak within the Island.
}

\begin{algorithm}
    \caption{DRUID class assignment algorithm}
    \label{algorithm:DRUID_Classes}
    \begin{algorithmic}
        \If{\texttt{component contains no children}}:
            \If{\texttt{has no parent}}:
                \State \texttt{Class = 0}    
            \Else{ \texttt{has parent}}: 
                \State \texttt{Class = 2}
            \EndIf 
            
        \Else{ \texttt{has children}}:
            \If{\texttt{has no parent}}:
                \State \texttt{Class = 4}
            \Else{ \texttt{has parent}}:
                \State\texttt{ Class = 3}
            \EndIf
        \EndIf
        
        \If{\texttt{created by CFD process}}:
            \State \texttt{Class = 1}
        \EndIf
        
    \end{algorithmic}
    \end{algorithm}
    This results in single component Islands being labelled Class 0, connected Islands labelled Class 4 and CFD points being Class 1, whilst Class 2 is the peak of features inside a connected Island (that is not the brightest) and Class 3 is an intermediate connected level. Figure \ref{fig:drawn_example_of_DRUID} shows how \new{components are} classified using this method.
    \item Source Characterisation - Further processing is done on the components to determine properties such as flux-weighted centroid coordinates, peak flux, flux density and others. A sample of the columns returned by the catalogue created by \DRUID\ are shown in Table \ref{tab:DRUID_outputs}.
    
\end{enumerate}

\subsubsection{Deblending}
\label{sec:deblending}
The result of persistent homology on an image returns a list of all 0-dimensional homology groups within the image that exist. This includes those that live for only a single pixel over a small intensity range. This full list of peaks and troughs in the image makes deblending a matter of filtering the groups we do not want. In \DRUID\ we have provided two parameters to control this. The first is the \texttt{lifetime\_limit\_fraction}, which is the minimum fractional difference between the peak and the trough of the group that destroys it. This can be thought of as a local significance of the group given its destruction level. The other option is an absolute value for this difference \texttt{lifetime\_limit}, it is good practice for this to be set as the standard deviation of the noise to remove pixel-to-pixel noise and spurious detections. These options provide flexibility to refine the level of deblending. Figure \ref{fig:drawn_example_of_DRUID} demonstrates how \DRUID\ will create different regions for an island of flux above \new{the} noise \new{with multiple components and one with just a single component}. The \new{Island on the left} will have \new{8} components, one for each of the peaks/births labelled as B \new{Class} N, where the N indicates the region Class. Class 3 regions show where merged peaks meet a Death (D) point. At the bottom, we have the Class 4 region drawn at the \texttt{analysis\_threshold}.

\subsubsection{Measuring Source Properties}

As previously mentioned, specific post-processing is done on the data depending on the image type. For the radio domain, we require specific characterisation steps to effectively measure component Class properties. Unresolved sources in radio data can be modelled as point sources convolved with a Gaussian\new{-}like beam and the pixel unit value is given in Jy/Beam. To recover the full flux density of a point source, we therefore must ensure we are accounting for the entire beam. This is simpler for Gaussian’s as the integrated flux density ($S$) is simply:

\begin{equation}
\label{eq:SGaus}
    S = S_{p} \frac{b_{maj}b_{min}}{\sigma_{maj} \sigma_{min}},
\end{equation}

Where $b_{maj}$ and $b_{min}$ are the major and minor axis of the beam and $\sigma_{min}$ and $\sigma_{maj}$ are the major and minor axis of the fitted Gaussian source. $S_{p}$ is the peak flux of the source measured in Jy/Beam. This is used by \pybdsf\ to calculate the flux densities of sources. However, as \DRUID\ does not assume a particula\new{r} morphology, we need to sum the pixel's flux densities per beam and apply the correct conversion. For a morphology agnostic calculation, we can follow the following relation.

\begin{equation}
    S = \sum_{i,j} S_{i,j} \frac{8 \ln{(2)} \delta_{pR} \delta_{pD}}{2 \pi b_{maj} b_{min}}.
\end{equation}
Where $\delta_{pR}$ and $\delta_{pD}$ are the angular sizes per pixel in the RA and DEC directions. For faint sources, where a significant amount of their emission is below the \texttt{analysis\_threshold}, we fail to sample enough of the beam which leads to an underestimation of the source Flux density. Source finding codes like \pybdsf\ \citep{mohan_pybdsf_2015} model the sources as Gaussian’s which solves this sampling problem automatically as equation \ref{eq:SGaus} would apply. \cite{hale_radio_2019} investigated the use of the optical source finder \profound\ on radio
data, and they found that it is essential to account for the under-sampling of this beam when calculating flux from the summation of pixels in a component region. To ensure we calculate the source flux density accurately we implemented a similar approach as is suggested in \cite{hale_radio_2019}. For each component, we create a model Gaussian beam generated from the BMIN, BMAJ and BPA parameters from the images restoring beam. We then calculate the flux density of this beam with the \new{source} component mask and without and
calculate the fractional difference, which is then multiplied by the measured flux density of the component. The corrected component flux $S_{c}$ is expressed as:

\begin{equation}
    S_{c} = S_n \frac{\sum_{i,j} B_{i,j}}{\sum_{i,j}M_{i,j}B_{i,j}},
\end{equation}

Where $S_{n}$ is the uncorrected component flux. $M$ is the component mask array and $B$ the model beam array. This correction factor is applied to components of all sizes but will only apply a negligible change to the flux contribution from bright and large sources. When the point spread function (PSF) of an optical image is larger than the \new{pixel size} of the image a similar problem of under-sampling the PSF occurs. \profound\ deals with this by dilating the source segmentation mask to include pixels below the threshold cut, ensuring full PSF sampling. SExtractor does this with a minimum-sized aperture that can be set to be larger than the images PSF. To correct for under-sampling we implement a similar technique to dealing with beam sampling in the radio regime, but though modelling the PSF. Table \ref{tab:source_findered_table} shows a summary of the main regime and methodology used by a selection of popular source finders, including the methods they use to define the source’s position.




\begin{table*}
\centering
\begin{tabular}{llllll}
\hline
Name                & Original Design Regime & Methodology                         & Source Position                              & Reference  \\ \hline 
\DRUID              & Radio \& Optical       & Persistent Homology                 & Brightest Pixel \& Flux weighted Centroid.   & This work. \\
\pybdsf             & Radio                  & Gaussian and wavelet Decomposition  & Gaussian Peak \& Flux weighted Centroid.     & \cite{mohan_pybdsf_2015}         \\
\profound           & Optical                & Watershed segmentation              & Flux weighted Centroid                       & \cite{robotham_profound_2018}                            \\
\texttt{SExtractor} & Optical                & Island Identification and apertures & Flux weighted Centroid among others       & \cite{bertin_sextractor_1996}                          \\
\texttt{AEGEAN}      & Radio                 & Gaussian Decomposition              & Gaussian Peak                                & \cite{hancock_source_2018,hancock_compact_2012}                              \\ \hline
\end{tabular}
\caption{Description of popular source finding methods. This includes \new{the} original design regime: (most have also been applied to others), the base methodology they utilise, the original reference and the main outputs from the source finders.}
\label{tab:source_findered_table}
\end{table*}

\subsection{\pybdsf}
\label{sec:pybdsf_explanation}
The Python Blob Detector and Source Finder\footnote{\href{https://github.com/lofar-astron/PyBDSF}{https://github.com/lofar-astron/PyBDSF}} (\pybdsf) was developed by \cite{mohan_pybdsf_2015}. \pybdsf\ has seen extensive use \new{for} LOFAR \new{images} and was used to generate the source catalogues in the LOFAR Two-Meter Sky Survey (LoTSS) data releases \cite{shimwell_lofar_2017, shimwell_lofar_2022}. \pybdsf\ is based on Gaussian decomposition to model and extract sources. This in principle will work well with point sources and smaller extended sources as the image has a Gaussian restoring beam. For more complex extended sources \pybdsf\ will model the emission by fitting multiple Gaussians. However, decomposing large-scale sources into Gaussian components will not always be effective. To improve this type of modelling, \pybdsf\ has included shapelet and wavelet decomposition into the source detection package. This is implemented after the initial Gaussian decomposition and the residual image is decomposed at various wavelet scales. In this work we will use \pybdsf\ with the wavelet decomposition functionality, \texttt{atrous\_do=True}. These catalogues however still required extensive manual inspection to look for spurious sources, make component associations and correct \pybdsf\ automatic source association \citep{hardcastle_lofar_2023, williams_lofar_2019}. \pybdsf\ has also been used by MIGHTEE \citep{heywood_mightee_2021} to create source catalogues, requiring the same post detection \citep{whittam_mightee_2024} inspections as LoTSS.

\subsection{\profound}
\label{sec:profound}

\profound\footnote{\href{https://github.com/asgr/ProFound}{https://github.com/asgr/ProFound}}
\citep{robotham_profound_2018} was designed to model galaxies by segmenting each object in the image. This uses pixel flux to make the segmentation tracing the emission without assuming the morphology of the object. A detailed description of how \profound\ works can be found in \cite{robotham_profound_2018}, it is integrated into the ProFit galaxy analysis codes. \cite{hale_radio_2019} showed that \profound\ can be useful in source extraction in radio images, particularly in extracting morphologically complex sources. The methodology \profound\ uses involves assigning pixels to segments \new{based} on whether they meet certain criteria. For example, a pixel neighbouring a segment with lower flux than the neighbour included in the segment will join it, so long as the flux \new{of the pixel} is above the specified flux limit. If a pixel has higher flux than its neighbours that are already assigned to a segment then it will not be assigned. The growing process terminates if there are no more pixels that can be assigned to the current segment. This watershed process fills out regions around pixels of interest. The deblending of sources by \profound\ is determined by the \texttt{tolerance} parameter, which controls how many sky RMS deviations to use to deblend sources during the watershed stage.




\section{Simulation of \new{Optical }Point Sources}
\label{sec:simulations}
\begin{figure}
    \centering
    \includegraphics[width=\linewidth]{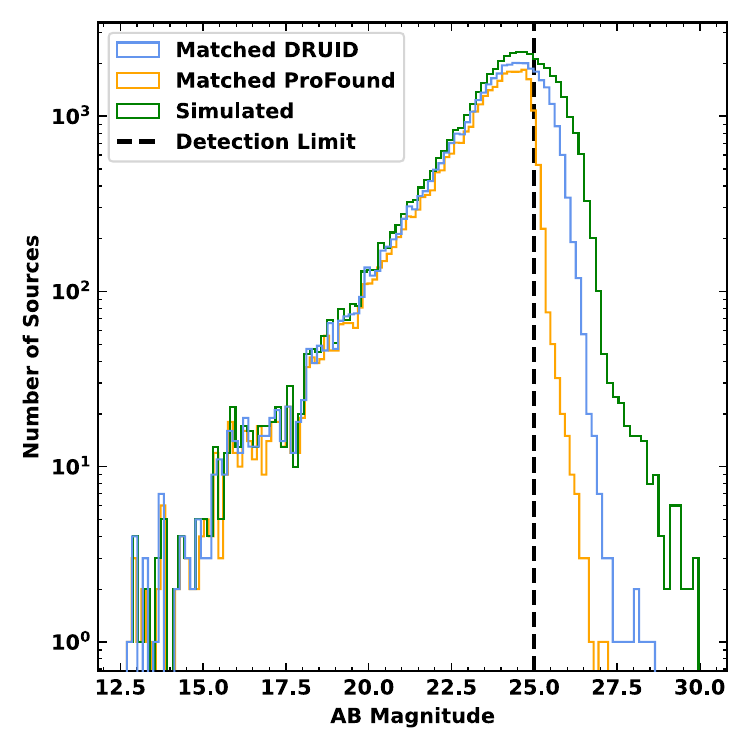}
    \caption{Distributions of AB Magnitudes from the injected sources \new{compared with} those recovered by \DRUID\ and \profound\ with the true injected simulations point sources.}
    \label{fig:sim_true_distribution}
\end{figure}

\begin{figure*}
    \centering
    \includegraphics[width=\linewidth]{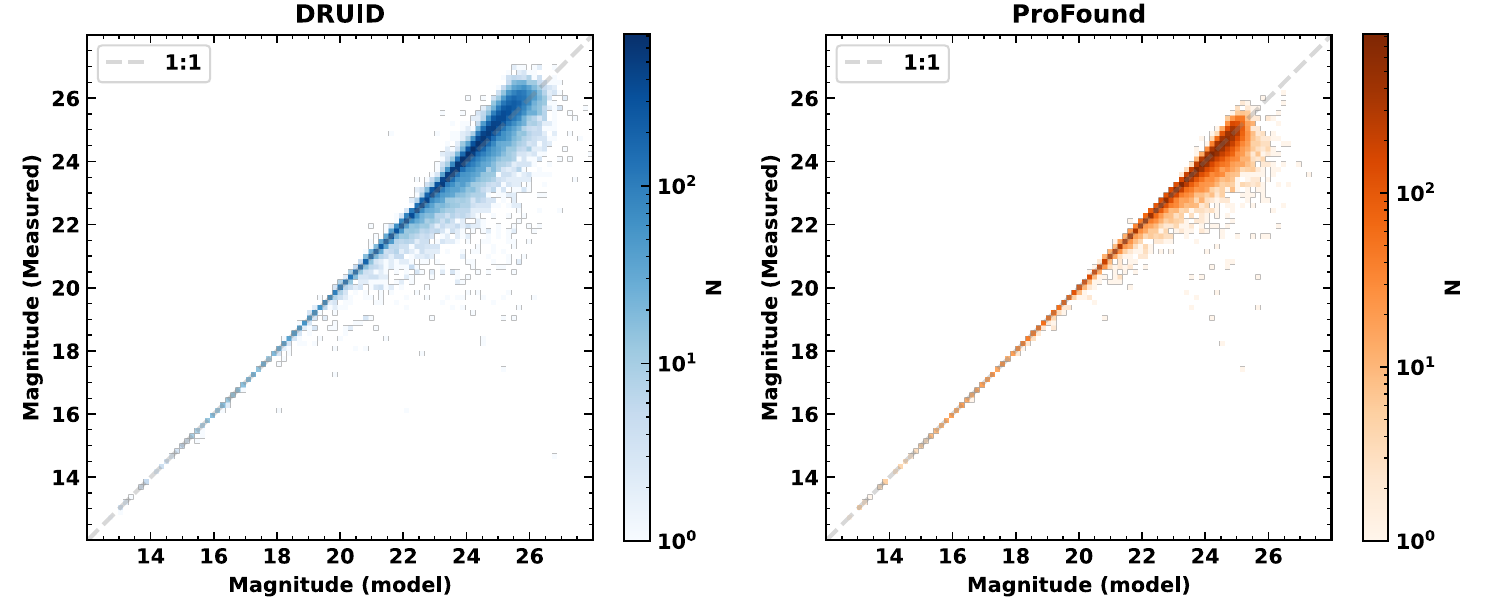}
    \caption{Recovered sources from the KiDS simulated point sources injection by \DRUID\ (left) and \profound\ (right). The matched sources are compared to the modelled magnitude from the truth catalogue.}
    \label{fig:Druid_ProFound_comparison_Magnitude}
\end{figure*}

\begin{figure}
    \centering
    \includegraphics[width=\linewidth]{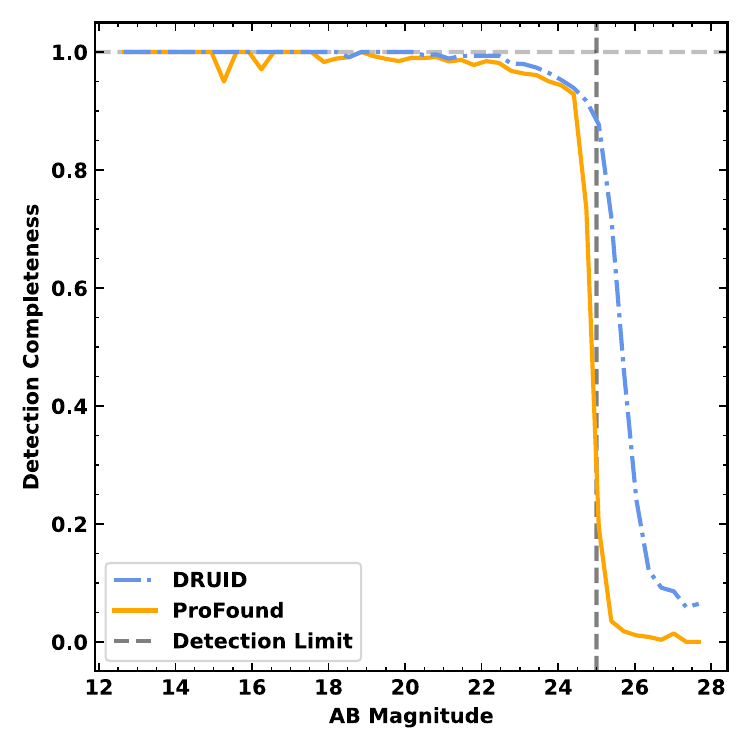}
    \caption{Detection completeness of \DRUID\ and \profound\ from the injection of simulated point sources into the example KiDS image. Sources are matched with a maximum separation of 0.5".}
    \label{fig:SIM_completeness}
\end{figure}

\begin{figure*}
    \centering
    \includegraphics[width=0.8\linewidth]{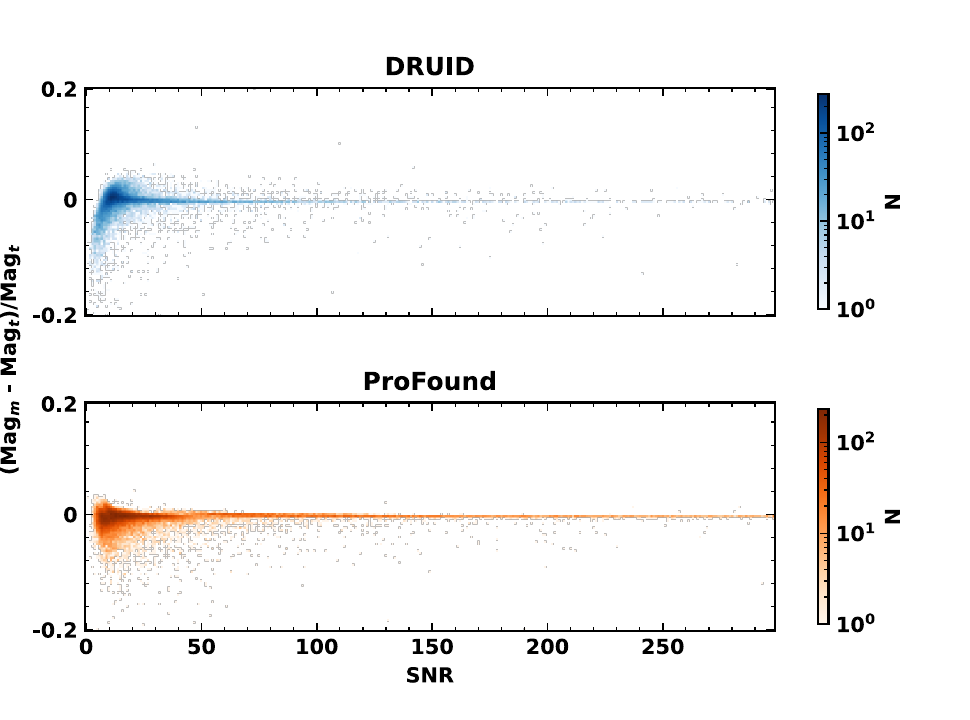}
    \caption{\DRUID\ (top) and \profound\ (bottom) relative magnitude difference with SNR compared to the truth catalogue.}
    \label{fig:mag_diffvssnr}
\end{figure*}

\subsection{Creating the Simulation}
To validate the effectiveness of \DRUID\ at recovering point sources in a real astronomical image we inject model sources into an example r-band KiDS image, specifically tile \texttt{KIDS\_20.4\_-34.1}. \new{This tile suffers from artefacts at the edges and saturating stars in the foreground so we remove a 1000 pixel border around the centre to remove most of these.} To create this simulated map of sources, we start by creating a distribution of source magnitudes taken from the DR4 source catalogue \citep{kuijken_fourth_2019}, we sample this distribution to gain a realistic sample of magnitudes to create the point sources \new{(}see Figure \ref{fig:sim_true_distribution}\new{)}. The location is then drawn at random from a uniform distribution, within the image dimensions. We then convert the magnitudes to fluxes and for each point source place a single point in that location. We then convolved this image with a Gaussian PSF with FWHM = 0.83", this is the PSF specified in the image header for this observation. All the point sources were then added to the \new{KiDS tile}. Due to the density of the image field, there is a significant chance of overlap or contamination by the other sources in the field. The resulting simulated catalogue contained 55,000 sources across the \new{reduced} image, at a resolution of 0.2" for \new{$\sim$0.9} deg$^{2}$ of sky coverage. The typical magnitude limit for the r-band \new{tiles} from the KiDS survey is 25, with a 2" diameter aperture for a $5\sigma$ detection \citep{de_jong_third_2017,kuijken_fourth_2019}

\subsection{Configuring Source finders}

We configure \DRUID\ to \new{preform source finding} with a detection threshold of $2\sigma$ with the same lower analysis threshold. Source characterisation is done in the "optical" mode. \DRUID\ also uses \texttt{cutup=True} with a buffer of 50 pixels. This is large enough to ensure that sources are fully detected in at least a single sub-image. We limit the components found by \DRUID\ to Class = 0, 1, 2 as these will best match the simple morphology of the sources. The full parameters and code used to configure \DRUID\ are shown in \ref{Appen:Druid_sim_code}. We also run \profound\ on the simulated image. We keep the parameters shared with \DRUID\ the same\new{,} these being \texttt{detection\_limit=2$\sigma$=SKYCUT}, \texttt{area\_limit=3=pixcut}, the parameters used can be found in \ref{Appen:Profound_sim_code}. Matching is performed by evaluating a simple sky separation distance where we match the closest corresponding source with a \new{maximum} separation distance of 0.5", which is approximately 2.5 pixels. This maximum separation was set to be inside the FWHM of the PSF, whilst allowing for some peak drift due to noise. For this \new{test with} \DRUID\ and \profound\ we use the location of the source peaks to calculate the sky coordinates, which better reflect the position of the Gaussian used in the simulated catalogue. This helps to ensure matching and reduce the effect of offset flux weighted positions if the source is injected into or near another bright source. The matching was performed using TOPCAT \citep{taylor_topcat_2005}.
 
\subsection{Simulation Results}

Figure \ref{fig:sim_true_distribution} shows the \new{magnitude} distribution of matched sources detected from \DRUID\ and \profound, we can see that the resulting distributions are very similar and suggest that despite \DRUID\ and \profound\ being set to the same detection threshold and same area rejection limit, \DRUID\ is detecting more sources \new{below 25 mag}. \new{There are} small variances in the \new{modelled} background distribution \new{between \DRUID\ and \profound, which could contribute to this difference.}

Figure \ref{fig:Druid_ProFound_comparison_Magnitude} shows matched magnitudes \new{of the sources} from \DRUID\ and \profound\ that have been matched to the truth catalogue used to create the injected point sources. We can see that both have a tight linear relationship between the detected and modelled magnitude as we approach the magnitude limit of 25. What is most notable is that \DRUID\ begins to underestimate some of the source's flux past a magnitude of 24. \profound\ on the other hand keeps a tighter scatter across the same region. This could be explained through the under-sampling of flux that is happening with \DRUID\ when approaching the detection limit. \DRUID\ does not allow for pixels lower than the lower analysis threshold to be included in the source region, resulting in a bias at low magnitude \new{due} to underestimating the source's total brightness. \profound\ avoids this problem with its \new{watershed} source dilation that allows for the inclusion of below the \texttt{SKYCUT} threshold. 
\new{The completeness of the detected sources from \DRUID\ and \profound\ is calculated as,}
\begin{equation}
    \new{Completeness = TP/(TP + FN),}
\end{equation}
\new{where $TP$ is the number of true positive sources detected and $FN$ is the number of undetected sources.} Figure \ref{fig:SIM_completeness} \new{shows} that \DRUID\ has completeness of \new{91\%} at a magnitude of 25 compared to \profound\ which has a completeness of \new{28\%}. \new{Contrary, \profound\ reaches 90\% completeness at 24.5 magnitude.} We have plotted the model flux from the matched sources, so the calculation of the sources magnitude by the source finder does not affect this completeness. \new{\DRUID\ has higher completeness at fainter magnitudes despite both source finders having similar limiting parameters such as \profound 's \texttt{pixcut} and \texttt{SKYCUT} parameters being set to the same as \DRUID 's equivalent parameters \texttt{detection\_threshold} and \texttt{area\_limit}}.

Figure \ref{fig:mag_diffvssnr} shows the measured magnitude difference compared to the injected catalogue with the SNR of that source. We see that \DRUID\ has some scatter at the lower SNR range, but this is comparable to the scatter in the same area of \profound . \new{\DRUID} tends to underestimate the magnitude of sources with low SNR. 
From manual inspection of the sources in the matched catalogues within the simulated image it was clear that several injected sources have been projected onto regions of extended emission. This means \DRUID\ is effectively deblending sources and finding these overlapping sources, but effectively modelling the flux from these sources is by its nature difficult.

\section{Application to Radio \new{Continuum Images}}

\subsection{RLAGN: Extended Radio Sources}
\label{sec:Analysis_3c}


\begin{figure*}
    \centering
    \includegraphics[width=\linewidth]{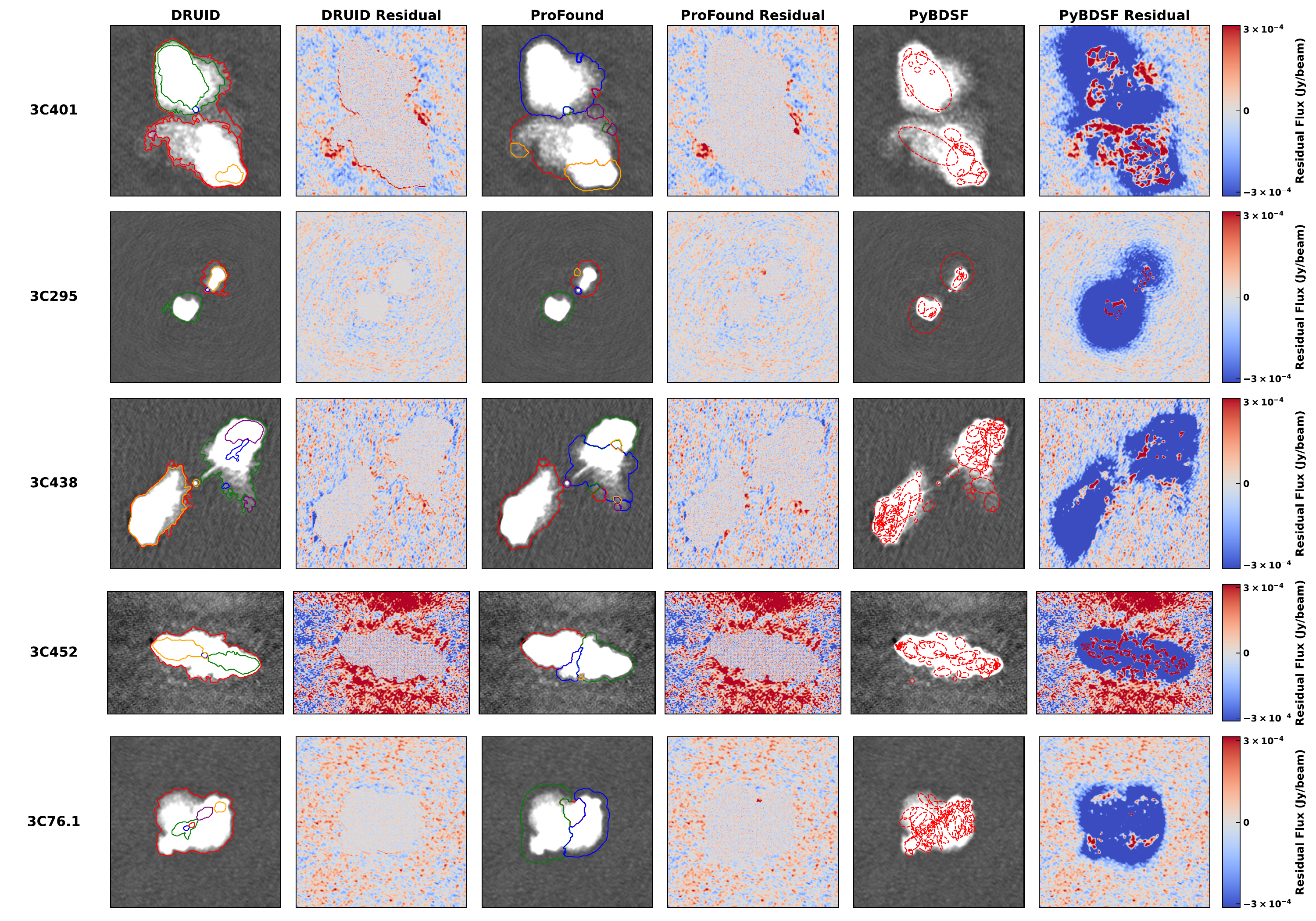}
    \caption{Mosaic of Radio sources 3C401, 3C295, 3C438, 3C452 and \new{3C76.1} (top to bottom) with the outcomes of \profound, \DRUID\ and \pybdsf\ source finding, showing source models and residuals. From left to right the columns are: image with \DRUID\ segmentation contours \new{overlaid}, \DRUID\ residual, image with \profound\ detection contours \new{overlaid}, \profound\ residual, \pybdsf\ fitted Gaussian ellipse \new{overlaid}, \pybdsf\ residual.}
    \label{fig:3CRR Mosaic}
\end{figure*}


\begin{figure}
    \centering
    \includegraphics[width=\linewidth]{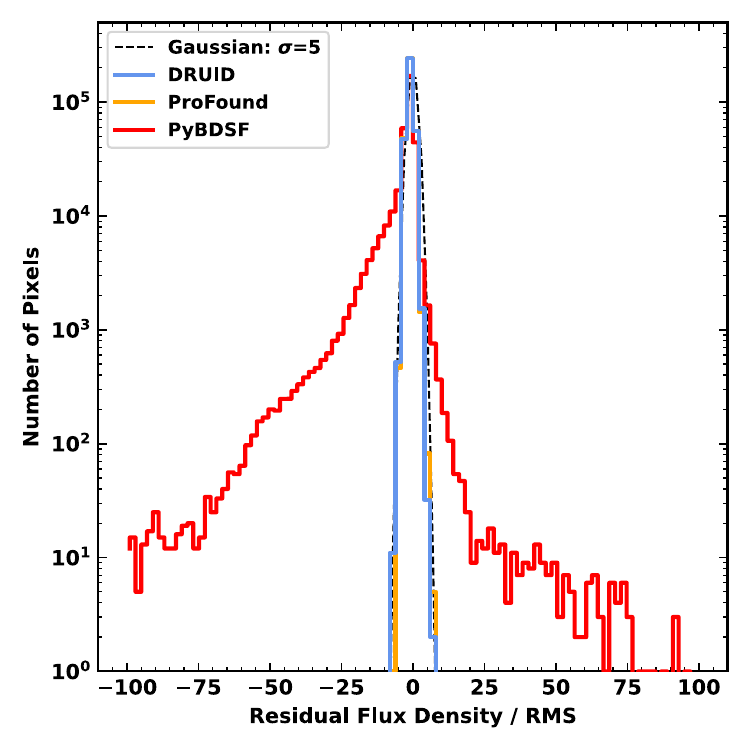}
    \caption{Residual distribution, \new{normalised with respect to local RMS,} from the 3CR radio sources described in \new{Table} \ref{tab:3C_AGN_params} and source detection shown in Figure \ref{fig:3CRR Mosaic}. Here the residual \new{distributions were} calculated for \DRUID, \pybdsf\ and \profound\ source finders. We also show a Gaussian distribution with \new{$\sigma=$ 5} \new{for comparison}.}
    \label{fig:3CR_noise_comparison}
\end{figure}

With the development of the next generation of radio surveys, \new{such as those to be performed with the SKA}, the automatic handling and analysis of detailed extended sources is becoming more important in the radio regime. Correctly identifying extended sources, being able to separate projected sources, and dealing with highly nested sources, is key to  making component associations and finding optical counterparts to RLAGN. To explore how \DRUID\ can segment a variety of morphologies, a group of RLAGN from the Third Cambridge catalogue \citep[3CR;][]{laing_bright_1983} was selected for segmentation. We compare \DRUID\ with \pybdsf\ and \profound, as both have seen use \new{for source finding in} radio \new{continuum data}. Figure \ref{fig:3CRR Mosaic} shows five 3CR RLAGN of both FRI and FRII morphology types and the resulting segmentation from \DRUID\ and \profound\ along with the Gaussian components from \pybdsf. \new{The radio data for these sources were} obtained from the Atlas of DRAGNs \footnote{\href{https://www.jb.man.ac.uk/atlas/alpha.html}{https://www.jb.man.ac.uk/atlas/alpha.html}}.

For each of the source finders, we have included a code snippet in Appendix \ref{appen:RLAGN_ALL}. Notably for \DRUID\ we used mode=’Radio’ and set the background \texttt{detection\_threshold=5} and \texttt{analysis\_threshold=2}. We set the same detection thresholds for \pybdsf\ and \profound\ and all have a minimum source size of 3 pixels. We also set the \texttt{lifetime\_limit\_fraction = 1.2}.  This was chosen as it is enough to show the main structures in the RLAGN that \DRUID\ can segment. \profound 's deblending parameter, \texttt{tolerance=15}, was given so that it can segment the main features from the RLAGNs as well. \pybdsf\ was configured with \texttt{thresh\_isl=5} and with \texttt{atrous\_do=True}\new{, allowing \pybdsf\ to use its own maximum wavelet scale}. Using the \texttt{atrous\_do} function is important as it models the residuals from the initial Gaussian decomposition into wavelets, an important step in attempting to better model extended emission. This function comes with a severe computational cost. The processing time for each of the RLAGN sources was $\sim$ 1 hour (on the computer described later in Section \ref{sec:comp_resources}), compared to the near instant runtime of \DRUID\ and \profound. \new{We use the} residual \new{created by} \pybdsf\ generated by subtracting the model Gaussian components from the image.

The segmentation results of each source finder are shown in the mosaic in Figure \ref{fig:3CRR Mosaic}. This shows the segmented regions for each of the five sources, \new{plotted on the original} image along with their residual image. All images are scaled consistently with each other and the residuals are also all to the same scale. We see that across the five RLAGN \DRUID\ and \profound\ both capture the morphology of the sources well. With \pybdsf, however, it is clear from the residual image that it does fail to correctly model the source flux, leading to over subtraction and gaps in coverage. This is still present despite using \pybdsf 's wavelet decomposition functionality. 
Comparing the internal components of the sources we can see how \DRUID\ \new{compares to \pybdsf\ and \profound\ in} segmenting and deblending features. We see the segmentation of jets, lobes, lobe bright spots and central bright spots. In comparison to \profound, whilst it also finds the same features, for the most part, the dilation from the watershed process pushes the boundaries of these features down towards the noise. This causes them to assign flux to the wrong component in some cases. This is particularly clear in 3C452, as the central peak has been given diffuse emission down towards the source boundary. This highlights an issue with \profound\ boundary dilation when segmenting features in complex images making the resulting segmentation oversized. In this case, where the RLAGN are in a low source density image, it does not pose a significant issue. \profound\ will struggle to deal with the complexities that come with nested sources, which will become an increasing problem in radio continuum imaging as new facilities start to come online. 

Figure \ref{fig:3CR_noise_comparison} shows the resulting \new{normalised} noise distribution from the combination of the 3CR sources in Figure \ref{fig:3CRR Mosaic}. These were generated from residuals of the respective source finders. If the source finder has extracted all the source associated flux from the image, we would expect a noise distribution that \new{resembles} a Gaussian distribution. A deviation from this distribution would suggest either over or under-fitting of the source. To generate these noise distributions, we combine the flattened residual images of the 3CR sources shown \new{and normalise them on their RMS} to form the histograms \new{shown in Figure \ref{fig:3CR_noise_comparison}}. 
This could introduce a bias in the resulting distribution, images with smaller sources as a fraction of the image size would contribute more to the noise distribution. Also, these images were taken at different frequencies and can have varying noise characteristics. However, as we are comparing the same images between each source finder this would affect all comparisons equally. It is clear from this Figure that \DRUID\ and \profound\ result \new{in near identical} residual noise distributions. This implies that despite the lack of dilation by \DRUID\ to include pixels below the noise threshold, we recover the same flux from the source. This means that this process has an insignificant effect on high SNR and morphologically complex sources. \pybdsf\ here fails to perform as well as the other methods as we see significant over-subtraction and under-subtraction of the sources. This \new{follows from} what we see in the residuals in Figure \ref{fig:3CRR Mosaic} and is consistent with the findings of \cite{hale_radio_2019}.

\begin{table*}
\centering
\begin{tabular}{lccccc}
\hline
Source & \multicolumn{1}{l}{Reference}  & \begin{tabular}[c]{@{}c@{}}Resolution \\ {[}arcsec{]}\end{tabular} & \begin{tabular}[c]{@{}c@{}}Frequency \\ {[}MHz{]}\end{tabular} & \begin{tabular}[c]{@{}c@{}}Size \\ {[}arcsec$^{2}${]}\end{tabular}\\ \hline
3C76.1 & \cite{leahy_vla_1991}    & 4.9                                                               & 1477                                                           & 383x384                                                         \\
3C295  & \cite{perley_vla_1991}    & 0.2                                                               & 8711                                                           & 15.4x15.4                                                          \\
3C452  & R.A. Laing (unpublished) & 6                                                                 & 1413                                                           & 317.2x345.8                                                       \\
3C438  & \cite[Leahy (1997),][]{treichel_spectral_2001}            & 0.29                                                              & 1534                                                           & 23.0x23.0                                                        \\
3C401  & J. P. Leahy (unplublished)  & 0.35                                                              & 1534   & 25.6x25.6    \\                                          
\hline
\end{tabular}
\caption{Information on 3C observations used in Section \ref{sec:Analysis_3c}. For each source, its reference, frequency, resolution and size. Data can be obtained from \protect\cite{leahy_atlas_2013} (\protect\href{https://www.jb.man.ac.uk/atlas/alpha.html}{www.jb.man.ac.uk/atlas/alpha.html}).}

\label{tab:3C_AGN_params}
\end{table*}




\subsection{LoFAR: Lockman Hole LoTSS Deep Field}
\label{sec:LoFarLockmanHole}

\new{The first public data release of the LoTSS Deep Fields \citep[][hereinafter refered to as T21]{tasse_lofar_2021} represents an opportunity to use \DRUID\ to create a source catalogue for one of the fields and compare it to the \pybdsf\ derived catalogue. In this section, we run DRUID on the LoTSS Deep Field image of the Lockman Hole (T21) and compare source matches to both the T21 \pybdsf\ radio source catalogue and the \cite[][hereinafter refered to as K21]{kondapally_lofar_2021} catalogue cross-matched to optical sources.} 

\new{
Following the T21 and K21 catalogues, we focus on the central 25 deg$^{2}$ of the observation which covers 30\% of the power of the primary beam. This observation is centred on 10h47m00s, +58d05m00s with a beam size of 6". The \pybdsf\ catalogue contains 50,112 sources, and the DRUID catalogue contains 56,758 sources from all Classes. We compare \DRUID\ to the \pybdsf\ source catalogue using \DRUID\ Classes 0, 1, 2 and 4. To compare the two catalogues we match them using TOPCAT \citep{taylor_topcat_2005} to a maximum separation of 3".
}

\begin{figure}
    \centering
    \includegraphics[width=\linewidth]{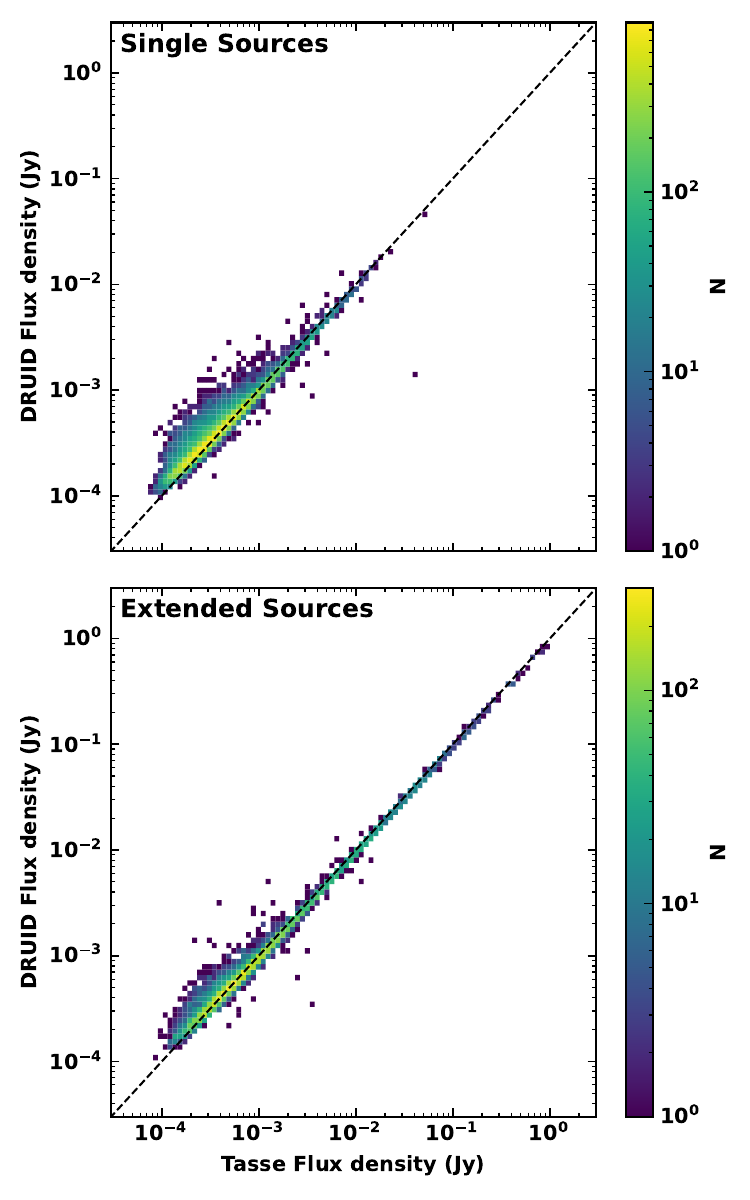}
    \caption{\new{Flux density comparison between \DRUID\ and \pybdsf s matched catalogues from the Lockman Hole LoTSS Deep Field \citep{tasse_lofar_2021}. The top plot shows only the single sources, and the bottom only extended  sources.}}
    \label{fig:DRUIDvspybdsf_tasseLoFAR}
\end{figure}

\new{
Figure \ref{fig:DRUIDvspybdsf_tasseLoFAR} shows a flux density comparison between the \pybdsf\ catalogue and \DRUID\ catalogues where sources have been matched. The top plot shows only the extended sources, and the bottom only single component sources. These are defined based on the number of Gaussians \pybdsf\ used to fit them, where extended is defined as anything requiring more than one Gaussian component. Both single sources and extended sources show strong agreement with each other, with some scatter and the occasional outlier. This trend is similar to what we see in Figure \ref{fig:Druid_ProFound_comparison_Magnitude}, where flux tends to be overestimated rather than underestimated. For bright extended sources, \DRUID\ begins to underestimate the flux density compared to \pybdsf. This effect is subtle but it is likely to be caused by the tendency of \pybdsf\ to overestimate the flux of the source, when dealing with complex extended sources as is shown in Figure \ref{fig:3CR_noise_comparison}.
}

\begin{figure}
    \centering
    \includegraphics[width=\linewidth]{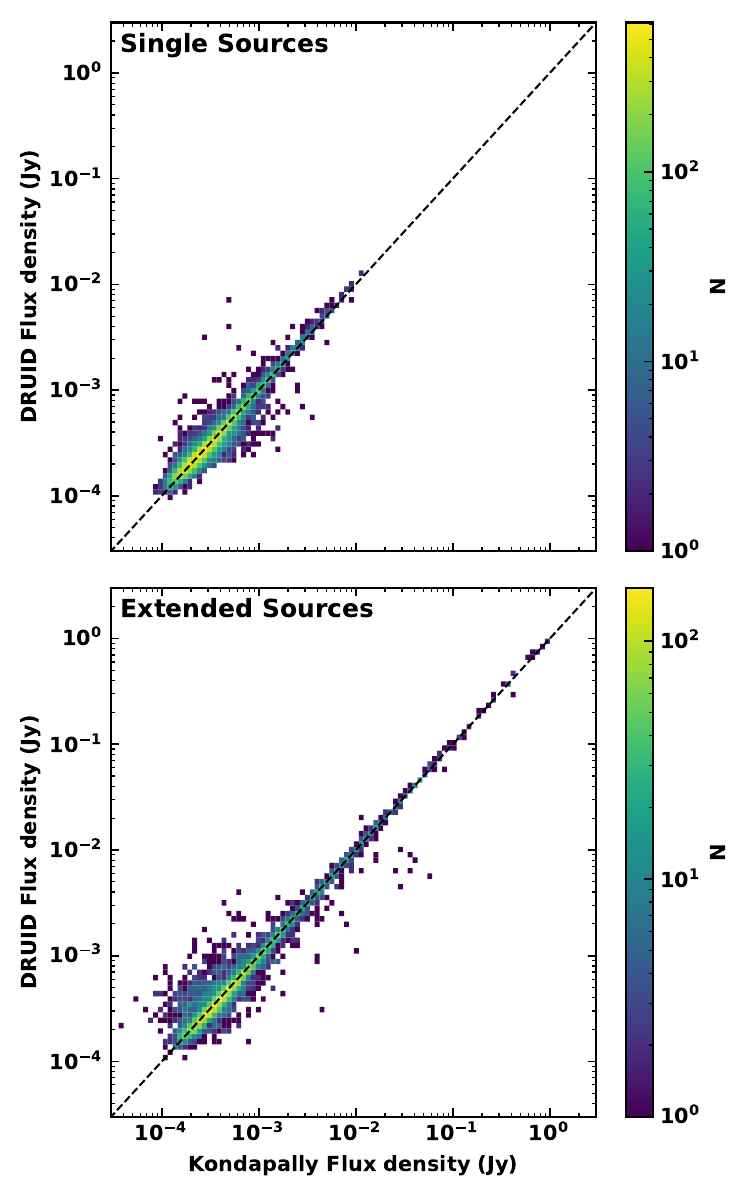}
    \caption{\new{Matched flux densities of the \DRUID\ LoTSS Deep Field Lockman Hole catalogue with the optical cross-matched catalogue \citep{kondapally_lofar_2021}. The top plot shows only the sources fitted with a single Gaussian by \pybdsf, while the lower plot shows sources that required more than one Gaussian.}}
    \label{fig:DRUId_kondapally_flux}
\end{figure}

\begin{figure*}
    \centering
    \includegraphics[width=\linewidth]{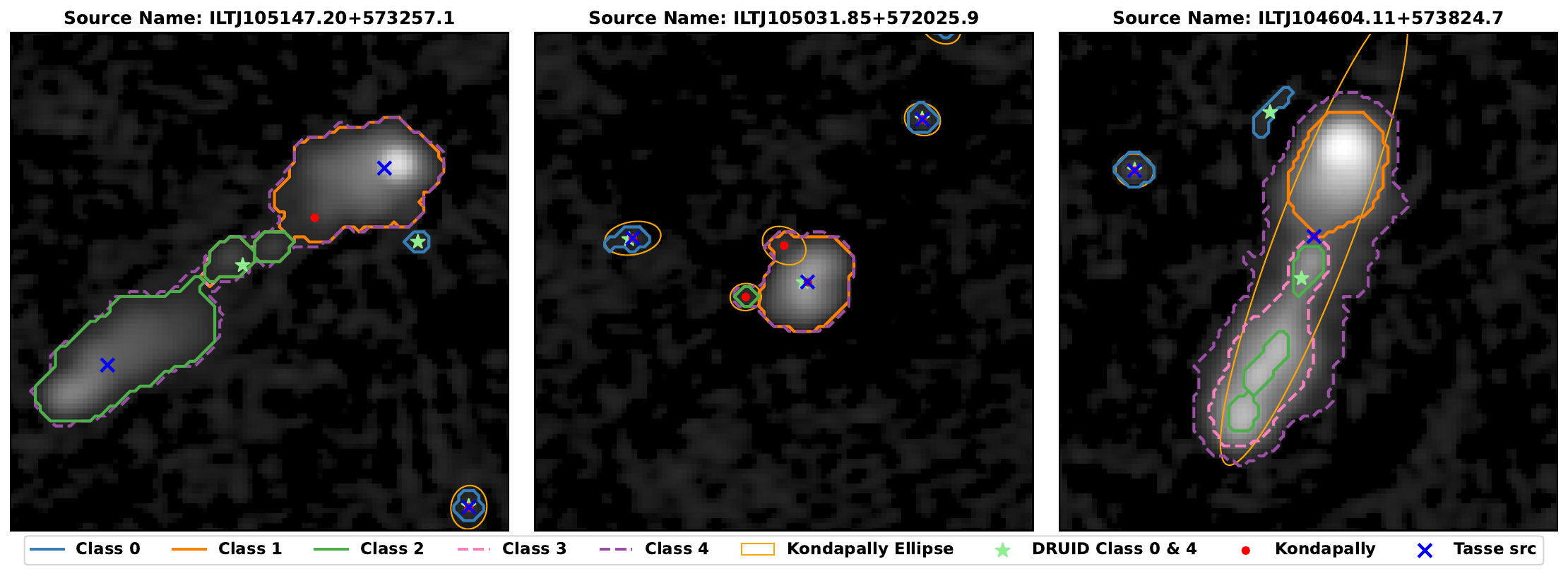}
    \caption{\new{Example of three sources within the LoTSS Lockman Hole Deep Field. Included are the DRUID Class regions, regions from the cross-matched \citep[][K21]{kondapally_lofar_2021} catalogue (yellow ellipse), the DRUID source location (green star), the K21 catalogue position (red filled circle) and the \pybdsf\ source \citep{tasse_lofar_2021} catalogue position (blue cross).}}
    \label{fig:DRUID_Tasse_Kondapally_images}
\end{figure*}

\new{
K21, along with cross-matching the sources from T21 with optical counterparts, allowed for deblending of confused sources and proper assignment of components either through a likelihood ratio or visual inspection.
We match the \DRUID\ catalogue to the cross-matched K21 catalogue with a maximum separation of 3" as for the T21 \pybdsf\ source catalogue. 
Figure \ref{fig:DRUId_kondapally_flux} shows the flux density comparison between \DRUID\ and the K21 catalogue. 
This shows a strong agreement between the fluxes measured by \DRUID\ and those of the cross-matched catalogue. 
The scatter in the plots is symmetrical around the 1-1 relation, contrary to the \pybdsf\ source catalogue in Figure \ref{fig:DRUIDvspybdsf_tasseLoFAR}, which shows \DRUID\ tends to overestimate the source flux when compared to \pybdsf. }

\new{The increase in outliers compared to Figure \ref{fig:DRUIDvspybdsf_tasseLoFAR} suggest that the \DRUID\ Class that is matching to the K21 catalogue does not always represent the same underlying source. 
We highlight the complexities in matching multi-component sources with the three examples in Figure \ref{fig:DRUID_Tasse_Kondapally_images}. 
We present three different sources from the field with the \DRUID\ Class regions, the fitted ellipse from the major and minor axes for the source if available and the marked position of the source in both \DRUID's catalogue, K21 catalogue and T21 catalogue.
The left panel shows that \DRUID\ has fitted this as a single Class 4 source, whilst in the T21 catalogue the two lobes were separated. 
The K21 catalogue contains this source, but no information on its altered size parameters. 
However, the flux value of \DRUID's Class 4 region is almost identical to the flux provided by K21. 
The position provided in K21 (red filled circle) is fairly offset from the flux weighted centroid of \DRUID's position (green star) and the two components in the T21 catalogue (blue crosses). 
The middle panel shows an example of a nearby source that \DRUID\ can separate and a confused one it cannot due to the low flux comparable to the noise level. 
The final panel shows how offset the source positions can be even when the T21 source has not been altered in the cross-matching process. This makes it difficult to ensure correct matching for larger extended objects.
}




\section{Application to Optical}
\subsection{Kilo-Degree Survey (KiDS)}
\label{sec:KiDs}

\begin{figure*}
    \centering
    \includegraphics[width=\linewidth]{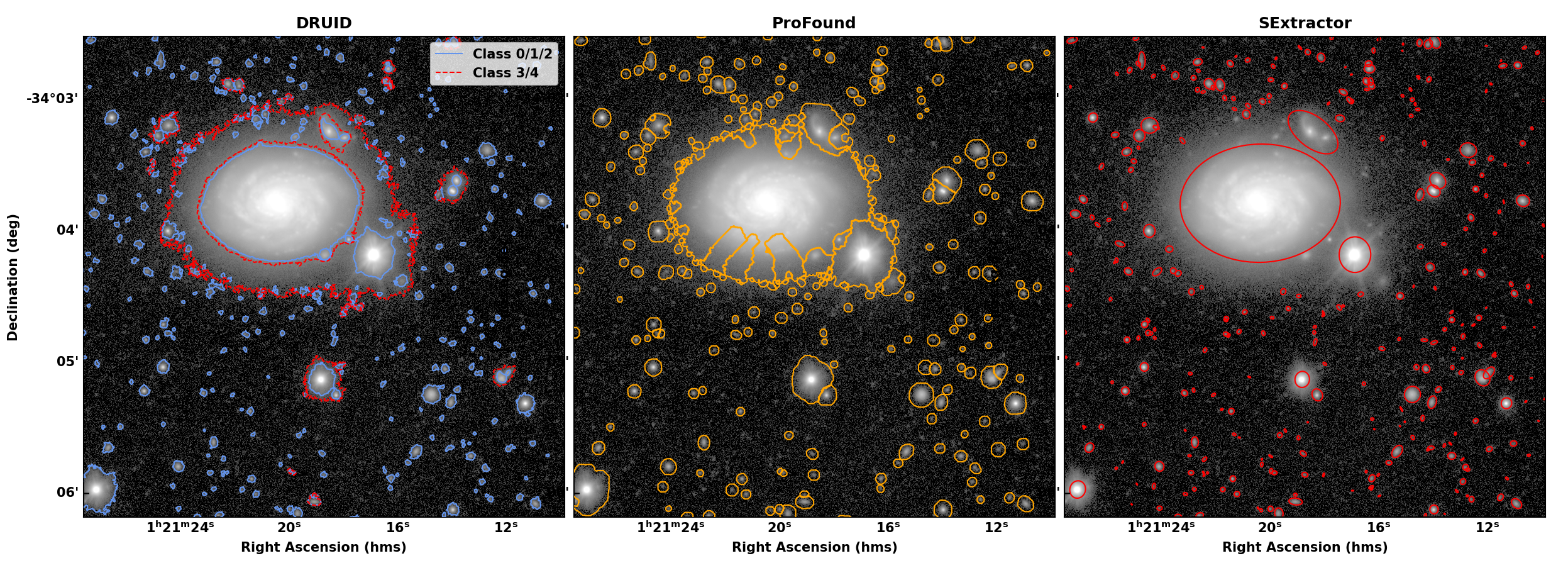}
    \caption{A cutout image from KiDS tile 20.4\_34.1 with source detection from \DRUID\ (left), \profound\ (center) and SExtractor (right) catalogues. The contours plotted show the regions used to extract flux by each source finder. With \DRUID\ we have distinguished the Class 0, 1, 2 from Class 3, 4. This is to show how Class choice changes the effective source boundary.}
    \label{fig:DRUID_KIDS_COMPARSION_IMAGE}
\end{figure*}

\begin{figure}
    \centering
    \includegraphics[width=\linewidth]{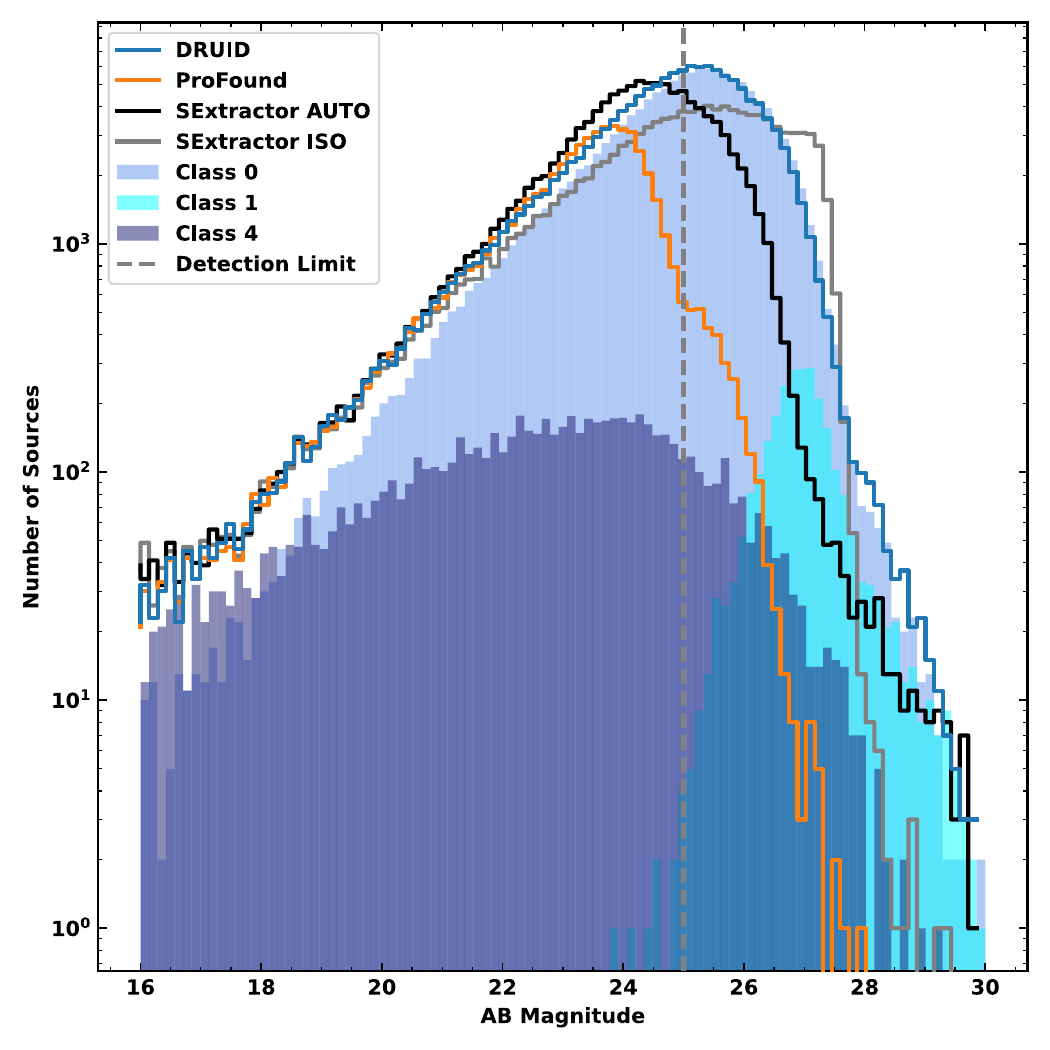}
    \caption{KiDS image distribution of sources detected by \DRUID\ and \profound\ compared to the KiDS DR3 SExtractor derived catalogue. Classes 0, 1 and 2 that make up the \DRUID\ distributions are also shown. The vertical dashed line represents the 5$\sigma$ detection limit.}
    \label{fig:KIDS_HIST_SOURCES}
\end{figure}

\begin{figure}
    \centering
    \includegraphics[width=\linewidth]{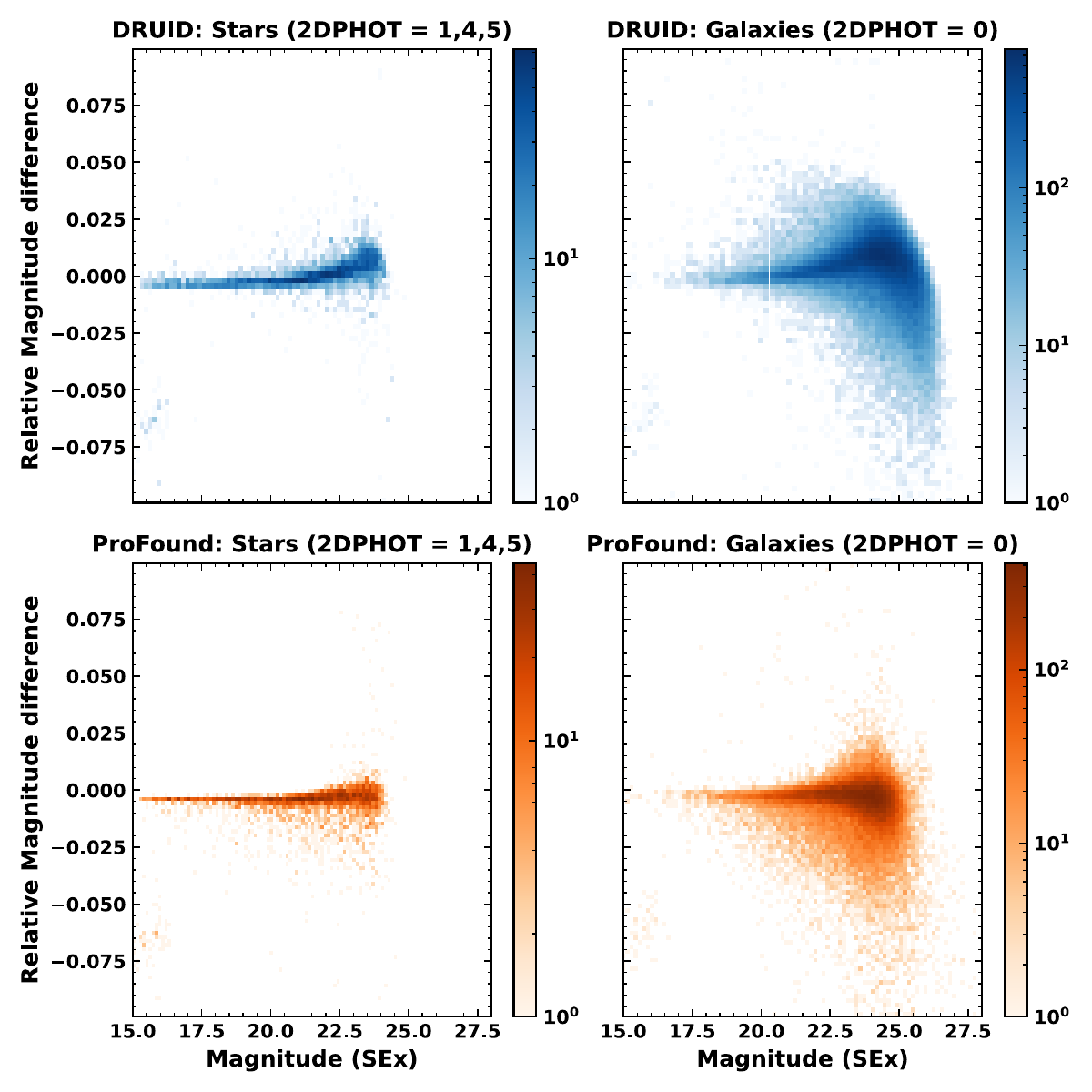}
    \caption{Relative difference between the SExtractor derived catalogue from the KiDS DR3 \citep{kuijken_fourth_2019} and the matched sources from \DRUID\ (top) and \profound\ (bottom) on the same image. These are separated in to object type Stars (left) and Galaxies (right).}
    \label{fig:MAG_Diff_StarGal}
\end{figure}

\begin{figure*}
    \centering
    \includegraphics[width=\linewidth]{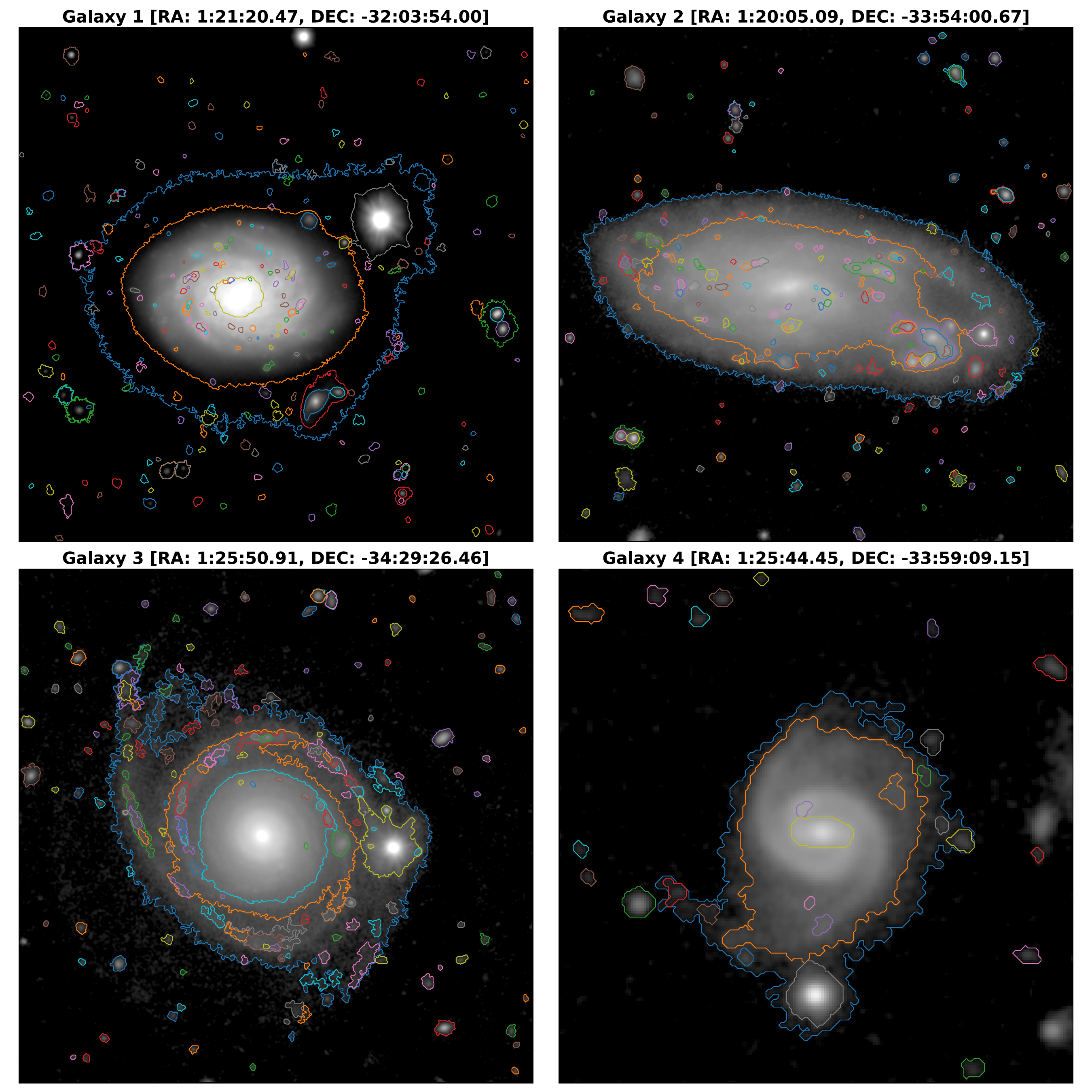}
    \caption{Examples of Galaxies from the KiDS tile 20.4\_\-34.1 and 21.6\_-34 with \DRUID\ detected segmented regions. This Figure shows the range of segmentation details that are extracted using \DRUID, detecting star forming regions within the galaxies in a way that still allows for the detection of the galaxy as a whole.}
    \label{fig:DRUID_KIDS_Galaxies}
\end{figure*}

The application of \DRUID\ also extends to the optical regime where we see much higher source density than within the radio regime. To explore this, we \new{perform} blind source detection with \DRUID\ and \profound\ on an image from the KiDS survey from the fourth data release \citep{kuijken_fourth_2019}. We analyse the r-band image centred on RA = 1h21m36s, Dec = -34d07m35s and KiDS tile ID: KIDS\_20.4\_-34.1. The associated catalogue was produced from the Astro-WISE three pipeline, which uses SExtractor for source extraction \citep{de_jong_third_2017}. This catalogue contains 118,068 sources which \new{are} made of a range of galaxies, stars, and image artefacts like diffraction spikes and saturated pixels.

\profound\ can be run on large images, but this comes with a memory penalty.  For \DRUID\ we had to ensure that we \new{created subimages}, with a defined buffer of 200 pixels. This is a necessary step as the full image size is \new{too large to be processed as a whole due to \texttt{cripser} limitations}. Whilst not necessary for completion, \DRUID\ was processed using GPU accelerated functions with \texttt{GPU = True}, this significantly improves the \new{time needed for} the source detection analysis. For all \new{codes}, we maintained a lower source \texttt{detection\_threshold} of 2 and \texttt{analysis\_threshold} of 2 and removed sources below an area of 5 pixels. The code detailing the full configurations of \DRUID\ and \profound\ can be found in the appendix Section \ref{appen:KIDS}. To achieve the best performance with \profound\ we adapted the \profound\ commands used by \cite{bellstedt_galaxy_2020}, who used \profound 's multi-band functionality to create photometric catalogues of KiDS/VIKINGS imaging. 

The resulting catalogues were matched to the DR4 catalogue with a maximum separation of 0.5" 
using TOPCAT. Since \DRUID\ can produce a some components with similar locations confusion can be present, choosing Classes= 0, 1 and 2 provided a close resemblance to the segmentation that would have been found by SExtractor and as seen in Figure \ref{fig:DRUID_KIDS_COMPARSION_IMAGE}. 
Figure \ref{fig:DRUID_KIDS_COMPARSION_IMAGE} demonstrates the differences between \DRUID, \profound, and SExtractor in segmenting the sources in this dense image. We can see that around the central resolved galaxy \DRUID\ and \profound\ can effectively find close sources that are absent from the SExtractor derived catalogue. \profound\ also seems to suffer from some edge effects around the \new{border} of bright sources. This is caused by artificially boosted noise by the flux island caused by the bright neighbour \new{leading to} spurious detections. \DRUID\ and SExtractor do not suffer from this. 

The source magnitude distributions of the final source catalogues from \DRUID , \profound, and SExtractor derived catalogues are shown in Figure \ref{fig:KIDS_HIST_SOURCES}. Here we also show the contribution of different \DRUID\ Classes. \DRUID\ detects a larger number of fainter sources in the image, from visual inspection of the source catalogue a number of these additional sources come from areas of raised background due to image artefacts. Whilst this is a problem it can be seen in Figure \ref{fig:DRUID_KIDS_COMPARSION_IMAGE} that \DRUID\ has deblended more sources from within bright galaxies and sources nearby to bright stars, which will cause an increase in the fainter end of the histogram and show as Class 2 sources. \DRUID\ performs much better at deblending nested sources, whilst limiting the number of spurious sources at their boundary. 

Figure \ref{fig:MAG_Diff_StarGal} shows the relative magnitude difference between \DRUID\ and \profound\ compared to SExtractor separated by source classification within the tile. We classify the sources using the 2DPHOT flag. This flag is derived from the results of the star/galaxy classification performed by SExtractor and rank based on confidence \citep{de_jong_first_2015}. We have defined as star any source with 2DPHOT=1,4,5 and galaxies \new{as sources with} 2DPHOT=0. We see that \DRUID\ and \profound\ perform similarly with both Classes up to a magnitude of 22. From here \DRUID\ tends to deviate from SExtractor's magnitude more than \profound. This is \new{similar to} the trend we observed with the injected point sources in Section \ref{sec:simulations}.

\subsection{Computation Resources}
\label{sec:comp_resources}

To access the utility of any source detection solution it is important to evaluate \new{their} computation demands to ensure the solution is scalable to large surveys. We evaluate the computational time, computer memory demands and GPU demands for \DRUID, \profound\ and SExtractor on the example KiDS tile described above in Section \ref{sec:KiDs}. These results are shown in Table \ref{tab:comp_res}. All codes were run on a machine with an AMD EPYC 7702P along with a Tesla T4 GPU. To evaluate SExtractor we used our run, with relevant parameters matching those of \DRUID\ and \profound. 
Due to how \DRUID\ has been implemented, a copy of the image exists in the GPU memory so the limit on GPU memory comes from the total size of the \new{sub-image} or image size. We see that \profound\ suffers from a significant memory penalty requiring nearly 10x more than \DRUID\ and SExtractor. As expected SExtractor is very fast and has very low memory requirements. Since \DRUID\ and \profound\ both deal with complex morphologies, they are inherently more computationally demanding than SExtractor. With large images, \DRUID\ requires a GPU to be competitive but requires far less memory than \profound.

\subsection{Resolved Galaxies}

To further demonstrate the use of \DRUID\ we examine how \DRUID\ segments complex morphological features from within resolved galaxies. These galaxies have been taken from the example KiDS tile in Section \ref{sec:KiDs} and tile KIDS\_21.6\_-34.1 and are shown in Figure \ref{fig:DRUID_KIDS_Galaxies}. To show the features within these galaxies we only require that the component has a lifetime that is more than background noise level. We see a variety of segmentation results possible with \DRUID. We see it can segment out star clumps within the galaxies and trace the present substructure. In these images, a neighbouring star has been effectively segmented and fainter neighbouring sources as well. These images show \DRUID s strengths at effectively segmenting features from within an image and reveals some substructure within the galaxy.

\begin{table}
\centering
\begin{tabular}{lllll}
\hline
Source Finder        & T$_{cpu}$ (s)& RAM (GB)& VRAM (GB)   &  Src/s   \\ \hline
\DRUID\ (GPU)          & \new{3,677}         & 8       & <1           &  \new{30}   \\
\DRUID\ (CPU)       & \new{11,932}        & 8       & -            &  \new{10}    \\
\profound             & 4,146         & 70      & -            &  23    \\
SExtractor           & 236          & 5       & -            &  1,171  \\ \hline
\end{tabular}
\caption{Summary of computation resources required for each source finder for blind source detection on KiDS\_0.4\_-34.1, with size 18708x20029 (1.4GB). Columns are; Total CPU time required, maximum memory used, maximum VRAM used and the sources per second that are processed.}
\label{tab:comp_res}
\end{table}

\section{Conclusion}

\DRUID\ detects and deblends sources using persistent homology utilising the creation and destruction of 0-dimensional homology groups. 
We have shown the utility of this method for deblending sources and that it can deal with the detailed morphologies of both RLAGN present in radio surveys and resolved galaxies from large optical surveys. 
\DRUID\ can effectively deal with a high level of nested sources without compromising the detection of neighbouring or extended sources in the same neighbourhood.
\DRUID\ can be used to generate source catalogues from large images and its computational demands make it reasonably scalable to processing larger surveys. \DRUID\ can be useful in a range of astronomical applications, due to its unique deblending and segmentation method.
To aid this \DRUID\ was designed to easily integrate custom functions between processing steps.
The success of \DRUID\ in being able to deal with highly nested sources meets a challenge as yet unmatched by another source detection and segmentation method.

\section*{Acknowledgements}
The authors would like to thank the reviewer for their comments and constructive feedback.
The authors would like to dedicate this work to Professor Mark Birkinshaw. Mark sadly passed away before the completion of this work. 
His contribution to this work was key in its early development and motivation. 
Mark's legacy lives on through the impact of his many published works and the many students he mentored and imparted some of his unique insights into astrophysics and life.
We would also like to thank Matthew Selwood, Fergus Baker and Thomas Higginson for their useful discussions during this work.
This work was supported by the UKRI Centre for Doctoral Training in Artificial Intelligence, Machine Learning \& Advanced Computing, funded by grant EP/S023992/1.
LOFAR data products were provided by the LOFAR Surveys Key Science project (LSKSP; https://lofar-surveys.org/) and were derived from observations with the International LOFAR Telescope (ILT). LOFAR \citep{van_haarlem_lofar_2013} is the Low Frequency Array designed and constructed by ASTRON. It has observing, data processing, and data storage facilities in several countries, which are owned by various parties (each with their own funding sources), and which are collectively operated by the ILT foundation under a joint scientific policy. The efforts of the LSKSP have benefited from funding from the European Research Council, NOVA, NWO, CNRS-INSU, the SURF Co-operative, the UK Science and Technology Funding Council and the Jülich Supercomputing Centre. 
Based on observations made with ESO Telescopes at the La Silla Paranal Observatory under programme IDs 177.A-3016, 177.A-3017, 177.A-3018 and 179.A-2004, and on data products produced by the KiDS consortium. The KiDS production team acknowledges support from: Deutsche Forschungsgemeinschaft, ERC, NOVA and NWO-M grants; Target; the University of Padova, and the University Federico II (Naples).


\section*{Data Availability}
Accompanying this work is the \DRUID\ source catalogue for the LOFAR LoTSS Deep Field Lockman Hole observation and tile KiDS\_20.4\_-34.1 of the KIDS survey. These and \DRUID\ can be found at \href{https://github.com/RhysAlfShaw/DRUID}{https://github.com/RhysAlfShaw/DRUID} and any bugs can be reported on the issues page.



\bibliographystyle{mnras}
\bibliography{references,additional_references} 

\appendix
\section{Simulation Code}
\subsection{\DRUID}
\label{Appen:Druid_sim_code}
\begin{verbatim}
findmysources = sf(image=image,image_path=None,
                mode='optical', cutup = True, 
                cutup_size = 1000, cutup_buff=100, 
                output = True, area_limit=3,
                smooth_sigma=1.5, nproc=1,
                GPU=True, header=header)
findmysources.set_background(detection_threshold=2,
                             analysis_threshold=2,
                             mode='mad_std')
findmysources.phsf(lifetime_limit_fraction=2)
findmysources.source_characterising(use_gpu=True)
\end{verbatim}  

\subsection{\profound}
\label{Appen:Profound_sim_code}

\begin{verbatim}
ProfoundSimulation=profoundProFound(image,
    skycut=2, reltol=-10,cliptol=100, pixcut=3,
    expandsigma=2,expand=1,roughpedestal = TRUE, 
    tolerance = 15,verbose=TRUE,rotstats=TRUE,
    boundstats=TRUE, nearstats=TRUE, 
    groupstats=TRUE, groupby='segim',
    deblend=TRUE)
\end{verbatim}

\section{RLAGN: Extended Radio Sources}
\label{appen:RLAGN_ALL}
\subsection{\DRUID}
\label{appen:RLAGN_DRUID}
\begin{verbatim}
    
findmysources = sf(image=image, image_path=None,
            mode='Radio', cutup = False, 
            cutup_size = None, cutup_buff=None,
            output = False, area_limit=3, 
            smooth_sigma=1, nproc=1,GPU=True, 
            header=None)
findmysources.set_background(detection_threshold=5,
                        analysis_threshold=2,
                        mode='Radio')
findmysources.phsf(lifetime_limit_fraction=1.2)
\end{verbatim}

\subsection{\pybdsf}
\label{appen:pybdsf_RLAGN}
\begin{verbatim}
    
img = bdsf.process_image(filename, atrous_do=True, 
            psf_snrcut=5.0, psf_snrcutstack=10.0,
            beam=(BMAJ, BMIN, beam_pa), 
            blank_limit=None, thresh='hard', 
            thresh_isl=5.0, thresh_pix=3.0,
            rms_map=True, rms_box=(70,70))
    
\end{verbatim}

\subsection{\profound}
\label{appen:RLAGN_profound}
\begin{verbatim}
Results = profoundProFound(image_matrix,
          plot=TRUE,verbose=TRUE,skycut=5,
          rotstats = TRUE, boundstats = TRUE,
          nearstats = TRUE, groupstats = TRUE,  
          smooth = FALSE, roughpedestal = TRUE,
          tolerance=15, cliptol=100,
          expandsigma=2, pixcut=3)
\end{verbatim}

\section{LOFAR: Lockman Hole}
\label{appen:LoFARLockman}
\subsection{\DRUID}
\label{appen:lofar_lockma_DRUID}
\begin{verbatim}
findmysources = sf(image=None, image_path=data_path,
        mode='Radio', cutup=True, cutup_size=1000, 
        cutup_buff=50,
        output=False, GPU=False)
findmysources.set_background(detection_threshold=5,
        analysis_threshold=2,
        bg_map_bool=True, box_size=50, 
        mode='mad_std')
findmysources.phsf(lifetime_limit_fraction=1.1)
findmysources.source_characterising(use_gpu=False)
\end{verbatim}

\section{Kilo-Degree Survey}
\label{appen:KIDS}
\subsection{\DRUID}
\label{appen:KIDSDRIUD}
\begin{verbatim}
findmysources = sf(image=image,image_path=None,
        mode='optical', cutup = True, 
        cutup_size = 1000, cutup_buff=200,
        output = False,
        area_limit=5, smooth_sigma=1.5,
        nproc=1, GPU=True, header=header)
findmysources.set_background(detection_threshold=2,
                     analysis_threshold=2,
                     mode='mad_std')
findmysources.phsf(lifetime_limit_fraction=2)
findmysources.source_characterising(use_gpu=True)
\end{verbatim}

\subsection{\profound}
\label{appen:KIDSprofound}
\begin{verbatim}
Results=profoundProFound(image,skycut=2, 
        reltol=-10,cliptol=100, pixcut=3,
        expandsigma=2,expand=1,roughpedestal = TRUE, 
        tolerance = 15,verbose=TRUE,
        rotstats=TRUE,boundstats=TRUE, 
        nearstats=TRUE, groupstats=TRUE, 
        groupby='segim',deblend=TRUE)
\end{verbatim}




\bsp	
\label{lastpage}
\end{document}